 \documentclass[preprint,review,12pt,times]{elsarticle}

\usepackage{caption}
\usepackage{amssymb}

\usepackage{titlesec}

\usepackage{lineno}

\usepackage[usenames,dvipsnames]{pstricks}
 \usepackage{epsfig}
 \usepackage{pst-grad} 
 \usepackage{pst-plot} 
 \usepackage[intlimits]{amsmath}
 \usepackage{tabularx}
 \usepackage{amssymb}
 \usepackage{setspace}
 \usepackage{epstopdf}
\usepackage{float}
\usepackage{fancyhdr}
 \usepackage{natbib}
 \biboptions{numbers,sort&compress}
 \usepackage{setspace}
 \usepackage{epstopdf}
 
\usepackage[T1]{fontenc}

\usepackage[latin1]{inputenc}

 \usepackage{tabularx}
 
\usepackage{caption}
\usepackage{footnote}
\usepackage{threeparttable}
\usepackage{graphicx}
\usepackage[caption=false]{subfig}
\usepackage{lipsum}

\journal{}

\begin{document}

\fancypagestyle{newstyle}{
\fancyhf{} 
\fancyhead[l]{\bfseries \nouppercase \rightmark} 
\fancyfoot[R]{\thepage} 
\renewcommand{\headrulewidth}{0pt}
\renewcommand{\footrulewidth}{0pt}}

\pagestyle{newstyle}

\begin{frontmatter}


\title{A Novel Design of Linear Phase Non-uniform Digital Filter Banks}


\author{Sakthivel V $^{\ast}$, Elizabeth Elias}

\address{Department of Electronics and Communication Engineering, National Institute of Technology Calicut, Kerala, India\\Email$^{\ast}$: sakthi517@nitc.ac.in, Tel: +91 9995335962, Fax: +91 4952287250 \\ \vspace{0.5cm}
}

\begin{abstract}

In many applications such as wireless communications and subband adaptive filtering, we need to design non-uniform filter banks (NUFB), which may lead to better performances and reduced hardware complexity when compared to uniform filter bank. NUFB satisfying linear phase property for all the constituent filters, are desirable in applications such as speech and image processing and in communication. This paper proposes a novel design of non-uniform modified discrete fourier transform filter bank (MDFT FB). Here, each non-uniform channel is obtained by merging the nearby channels of a uniform MDFT FB.  The reported works of non-uniform cosine modulated filter banks (CMFBs) do not satisfy linear phase property for all the constituent filters. In this work, we introduce the design of a non-uniform MDFT FB which satisfies linear phase property for all the constituent filters. The proposed design of the non-uniform MDFT FB is checked utilizing the alias cancellation among the channels, distortion and flatness condition of the channels. A condition is derived to find the channels when merged with adjacent channels, will cause aliasing in M channels. In the design and implementation of the proposed non-uniform MDFT FB, the structure of the uniform MDFT filter bank is preserved and hence all the advantages of MDFT FB over the DFT filter bank, are guaranteed. Hence without increasing the design complexity, the non-uniform MDFT FB is designed.  
\end{abstract}


\end{frontmatter}


\section{Introduction} 
                                                                                                                                                                                                                        \label{intro}                                                                                                                                                                                                                                                                                                                                   Filter banks (FB) are used in many applications such as wireless communication, image compression, speech processing, sub-band coding and adaptive signal processing \citep{pp}. Due to the simple design and easy realization, modulated FB are more popular than other FB structures. Mainly there are two types of modulated FB, Discrete Fourier Transform (DFT) FB and Cosine modulated filter bank (CMFB). Using the same prototype filter, the analysis and synthesis FB are generated by modulation of the prototype filter. The resultant DFT FB have analysis and synthesis filters which satisfy linear phase property if the prototype filter is chosen to have linear phase. CMFB has only overall linear phase. Hence DFT FB are preferred in many applications. However, DFT FB have no inherent alias cancellation structure. 

\par
 To overcome this disadvantage of DFT FB, modified DFT filter bank (MDFT FB) can be used \citep{fliege1993computational,fliege1994modified,sakthivel2014design,sakthivel2015design}. MDFT FB provide linear phase in both the analysis and synthesis filters, provided, the prototype filter is chosen to have linear phase. In MDFT FB, alias cancellation is available in the structure, which will automatically cancel all the odd alias spectra. This leads to near perfect reconstruction (NPR) MDFT FB. Perfect reconstruction (PR) condition on FB, which lead to zero error at the output, may result in more complex FB than is actually required to meet the specifications. Design and hardware complexity can be reduced by using NPR instead of PR FB. Hence for many practical applications, FB satisfying the NPR property are better choices as long as the distortions are within the limits specified by the applications \citep{bregovic2002efficient}. In NPR MDFT FB, the aliasing distortion and amplitude distortion are very low \citep{fliege1993computational,fliege1994modified,sakthivel2014design,sakthivel2015design}.

\par 

In many applications such as wireless communications and subband adaptive filtering , we need to design non-uniform filter banks (NUFB) \citep{griesbach1999non,li2006efficient,jun2010efficient,nongpiur2012maximizing}. NUFB, may lead to better performances and reduced complexity. There are some works already available on the design of NUFB. In \citep{hoang1989non}, a non-uniform quadrature mirror FB is designed particularly for integer valued decimation factors. In \citep{nayebi1993nonuniform}, a NUFB is designed directly by the time domain analysis and in \citep{lee1995design}, a non-uniform CMFB is designed by merging the adjacent channels. However, in CMFB,  the individual filters do not have linear phase. In this work, we introduce the design of a non-uniform NPR MDFT FB which is obtained from a NPR uniform MDFT FB. All the desirable characteristics of uniform MDFT FB such as linear phase of all constituent filters and very low amplitude distortion and aliasing are retained in the proposed non-uniform MDFT FB. Here every constituent non-uniform channel is obtained by merging some of the appropriate adjacent filters in the uniform MDFT FB. Hence all the preferred attributes of the uniform MDFT FB are safeguarded and the design procedures are similar to those of the uniform MDFT FB \citep{fliege1994modified}. This design of non-uniform MDFT FB by merging adjacent channels is not proposed in the literature so far.

\par 

The rest of the paper is organized as follows: Section 2 gives an introduction of MDFT uniform FB. Section 3 proposes the design of non-uniform MDFT FB. Section 4 gives the details of results and analysis. Section 5 concludes the paper. 

\label{sec:1}

\section{MDFT uniform FB }

In the structure of the DFT FB \citep{fliege1994multirate,sakthivel2014design}, $H_{k}(z)$ and $F_{k}(z)$ represent the analysis and synthesis filters respectively of an M-channel FB and $\uparrow M$ and $\downarrow M$ represent interpolation and decimation by M respectively as shown in Fig.1. Here, the analysis and synthesis filters are derived by complex modulation from a linear phase finite impulse response (FIR) prototype filter $H(z)$ with a transition band from $-\pi/M$ to $\pi/M$.

\begin{figure}[here]
\centering

\psscalebox{0.7 0.7} 
{
\begin{pspicture}(0,-5.0)(18.78,5.0)
\psframe[linecolor=black, linewidth=0.04, dimen=outer](3.6,3.8)(2.0,2.2)
\psframe[linecolor=black, linewidth=0.04, dimen=outer](7.2,3.8)(5.6,2.2)
\psframe[linecolor=black, linewidth=0.04, dimen=outer](12.8,3.8)(11.2,2.2)
\psframe[linecolor=black, linewidth=0.04, dimen=outer](3.6,1.4)(2.0,-0.2)
\psframe[linecolor=black, linewidth=0.04, dimen=outer](7.2,1.4)(5.6,-0.2)
\psframe[linecolor=black, linewidth=0.04, dimen=outer](7.2,-2.6)(5.6,-4.2)
\psframe[linecolor=black, linewidth=0.04, dimen=outer](12.8,-2.6)(11.2,-4.2)
\psline[linecolor=black, linewidth=0.04, arrowsize=0.05291666666666668cm 2.0,arrowlength=1.4,arrowinset=0.0]{->}(3.6,3.0)(5.6,3.0)
\psline[linecolor=black, linewidth=0.04, arrowsize=0.05291666666666668cm 2.0,arrowlength=1.4,arrowinset=0.0]{->}(3.6,0.6)(5.6,0.6)
\psline[linecolor=black, linewidth=0.04, arrowsize=0.05291666666666668cm 2.0,arrowlength=1.4,arrowinset=0.0]{->}(3.6,-3.4)(5.6,-3.4)
\psline[linecolor=black, linewidth=0.04, arrowsize=0.05291666666666668cm 2.0,arrowlength=1.4,arrowinset=0.0]{->}(10.0,0.6)(11.2,0.6)
\psframe[linecolor=black, linewidth=0.04, dimen=outer](16.4,3.8)(14.8,2.2)
\psframe[linecolor=black, linewidth=0.04, dimen=outer](16.4,1.4)(14.8,-0.2)
\psline[linecolor=black, linewidth=0.04, arrowsize=0.05291666666666668cm 2.0,arrowlength=1.4,arrowinset=0.0]{->}(12.8,3.0)(14.8,3.0)
\psframe[linecolor=black, linewidth=0.04, dimen=outer](12.8,1.4)(11.2,-0.2)
\psline[linecolor=black, linewidth=0.04, arrowsize=0.05291666666666668cm 2.0,arrowlength=1.4,arrowinset=0.0]{->}(12.8,0.6)(14.8,0.6)
\psline[linecolor=black, linewidth=0.04, arrowsize=0.05291666666666668cm 2.0,arrowlength=1.4,arrowinset=0.0]{->}(12.8,-3.4)(14.8,-3.4)
\rput[bl](6.0,3.0){$\downarrow M$}
\rput[bl](15.2,3.0){$F_{0}(z)$}
\rput[bl](11.6,3.0){$\uparrow M$}
\rput[bl](2.4,0.6){$H_{1}(z)$}
\rput[bl](6.0,0.6){$\downarrow M$}
\rput[bl](2.4,3.0){$H_{0}(z)$}
\rput[bl](11.6,0.6){$\uparrow M$}
\rput[bl](15.2,0.6){$F_{1}(z)$}
\rput[bl](6.0,-3.4){$\downarrow M$}
\rput[bl](11.6,-3.4){$\uparrow M$}
\psline[linecolor=black, linewidth=0.04, arrowsize=0.05291666666666668cm 2.0,arrowlength=1.4,arrowinset=0.0]{->}(0.8,3.0)(0.8,3.0)(2.0,3.0)
\psline[linecolor=black, linewidth=0.04, arrowsize=0.05291666666666668cm 2.0,arrowlength=1.4,arrowinset=0.0]{->}(0.8,0.6)(0.8,0.6)(2.0,0.6)
\psline[linecolor=black, linewidth=0.04, arrowsize=0.05291666666666668cm 2.0,arrowlength=1.4,arrowinset=0.0]{->}(0.8,4.2)(0.8,0.6)
\psline[linecolor=black, linewidth=0.04, arrowsize=0.05291666666666668cm 2.0,arrowlength=1.4,arrowinset=0.0]{->}(0.8,0.6)(0.8,0.6)(0.8,-0.6)
\psline[linecolor=black, linewidth=0.04, arrowsize=0.05291666666666668cm 2.0,arrowlength=1.4,arrowinset=0.0]{->}(0.8,-3.4)(2.0,-3.4)
\psline[linecolor=black, linewidth=0.04, arrowsize=0.05291666666666668cm 2.0,arrowlength=1.4,arrowinset=0.0]{->}(0.8,-2.2)(0.8,-3.4)
\psline[linecolor=black, linewidth=0.04, arrowsize=0.05291666666666668cm 2.0,arrowlength=1.4,arrowinset=0.0]{->}(16.4,3.0)(17.6,3.0)
\psline[linecolor=black, linewidth=0.04, arrowsize=0.05291666666666668cm 2.0,arrowlength=1.4,arrowinset=0.0]{->}(16.4,0.6)(17.6,0.6)
\psline[linecolor=black, linewidth=0.04, arrowsize=0.05291666666666668cm 2.0,arrowlength=1.4,arrowinset=0.0]{->}(16.4,-3.4)(17.6,-3.4)
\psline[linecolor=black, linewidth=0.04, arrowsize=0.05291666666666668cm 2.0,arrowlength=1.4,arrowinset=0.0]{->}(17.6,3.0)(17.6,0.6)
\psline[linecolor=black, linewidth=0.04, arrowsize=0.05291666666666668cm 2.0,arrowlength=1.4,arrowinset=0.0]{->}(17.6,0.6)(17.6,-0.6)
\psline[linecolor=black, linewidth=0.04, arrowsize=0.05291666666666668cm 2.0,arrowlength=1.4,arrowinset=0.0]{->}(17.6,-2.6)(17.6,-4.6)

\rput[bl](6.4,-0.2){.}
\psdots[linecolor=black, dotsize=0.06](2.8,-0.2)
\psdots[linecolor=black, dotsize=0.06](2.8,-0.2)
\psdots[linecolor=black, dotsize=0.06](2.8,-0.2)
\psdots[linecolor=black, dotsize=0.06](2.8,-0.2)
\psdots[linecolor=black, dotsize=0.06](2.8,-0.6)
\psdots[linecolor=black, dotsize=0.06](2.8,-0.6)
\psdots[linecolor=black, dotsize=0.06](2.8,-1.0)
\psdots[linecolor=black, dotsize=0.06](2.8,-1.4)
\psdots[linecolor=black, dotsize=0.06](2.8,-1.8)
\psdots[linecolor=black, dotsize=0.06](2.8,-2.2)
\psdots[linecolor=black, dotsize=0.06](2.8,-0.6)
\psdots[linecolor=black, dotsize=0.06](2.8,-0.6)
\psdots[linecolor=black, dotsize=0.06](2.8,-2.6)
\psdots[linecolor=black, dotsize=0.06](0.8,-0.6)
\psdots[linecolor=black, dotsize=0.06](0.8,-0.6)
\psdots[linecolor=black, dotsize=0.06](0.8,-1.0)
\psdots[linecolor=black, dotsize=0.06](0.8,-1.4)
\psdots[linecolor=black, dotsize=0.06](0.8,-1.8)
\psdots[linecolor=black, dotsize=0.06](6.4,-0.2)
\psdots[linecolor=black, dotsize=0.06](6.4,-0.6)
\psdots[linecolor=black, dotsize=0.06](6.4,-1.0)
\psdots[linecolor=black, dotsize=0.06](6.4,-1.4)
\psdots[linecolor=black, dotsize=0.06](6.4,-1.8)
\psdots[linecolor=black, dotsize=0.06](6.4,-2.2)
\psdots[linecolor=black, dotsize=0.06](6.4,-2.6)
\psdots[linecolor=black, dotsize=0.06](12.0,-0.2)
\psdots[linecolor=black, dotsize=0.06](12.0,-0.6)
\psdots[linecolor=black, dotsize=0.06](12.0,-1.0)
\psdots[linecolor=black, dotsize=0.06](12.0,-1.4)
\psdots[linecolor=black, dotsize=0.06](12.0,-1.8)
\psdots[linecolor=black, dotsize=0.06](12.0,-2.2)
\psdots[linecolor=black, dotsize=0.06](12.0,-2.6)
\psdots[linecolor=black, dotsize=0.06](15.6,-0.2)
\psdots[linecolor=black, dotsize=0.06](15.6,-0.6)
\psdots[linecolor=black, dotsize=0.06](15.6,-1.0)
\psdots[linecolor=black, dotsize=0.06](15.6,-1.4)
\psdots[linecolor=black, dotsize=0.06](15.6,-1.8)
\psdots[linecolor=black, dotsize=0.06](15.6,-2.2)
\psdots[linecolor=black, dotsize=0.06](15.6,-2.6)
\psdots[linecolor=black, dotsize=0.06](17.6,-1.0)
\psdots[linecolor=black, dotsize=0.06](17.6,-1.4)
\psdots[linecolor=black, dotsize=0.06](17.6,-1.8)
\psdots[linecolor=black, dotsize=0.06](17.6,-2.2)
\psdots[linecolor=black, dotsize=0.06](17.6,-2.6)
\psline[linecolor=black, linewidth=0.04, arrowsize=0.05291666666666668cm 2.0,arrowlength=1.4,arrowinset=0.0]{->}(10.4,-3.4)(11.2,-3.4)
\psline[linecolor=black, linewidth=0.04, arrowsize=0.05291666666666668cm 2.0,arrowlength=1.4,arrowinset=0.0]{->}(10.0,3.0)(11.2,3.0)
\psframe[linecolor=black, linewidth=0.04, dimen=outer](3.6,-2.6)(2.0,-4.2)
\psframe[linecolor=black, linewidth=0.04, dimen=outer](16.4,-2.6)(14.8,-4.2)
\rput[bl](2.0,-3.4){$H_{M-1}(z)$}
\rput[bl](14.8,-3.4){$F_{M-1}(z)$}
\rput[bl](0.0,4.2){$x(n)$}
\rput[bl](17.6,-5.0){$\hat{x}(n)$}
\rput[bl](7.6,3.4){$y_{0}(m)$}
\rput[bl](9.6,3.4){$\hat{y}_{0}(m)$}
\rput[bl](7.6,1.0){$y_{1}(m)$}
\rput[bl](9.6,1.0){$\hat{y}_{1}(m)$}
\rput[bl](7.6,-3.0){$y_{M-1}(m)$}
\rput[bl](9.6,-3.0){$\hat{y}_{M-1}(m)$}
\psline[linecolor=black, linewidth=0.04, arrowsize=0.05291666666666667cm 2.0,arrowlength=1.4,arrowinset=0.0]{->}(7.2,3.0)(8.4,3.0)
\psline[linecolor=black, linewidth=0.04, arrowsize=0.05291666666666667cm 2.0,arrowlength=1.4,arrowinset=0.0]{->}(7.2,0.6)(8.4,0.6)
\psline[linecolor=black, linewidth=0.04, arrowsize=0.05291666666666667cm 2.0,arrowlength=1.4,arrowinset=0.0]{->}(7.2,-3.4)(8.4,-3.4)
\end{pspicture}
}

\caption{DFT filter bank}
\label{dft}

\end{figure}
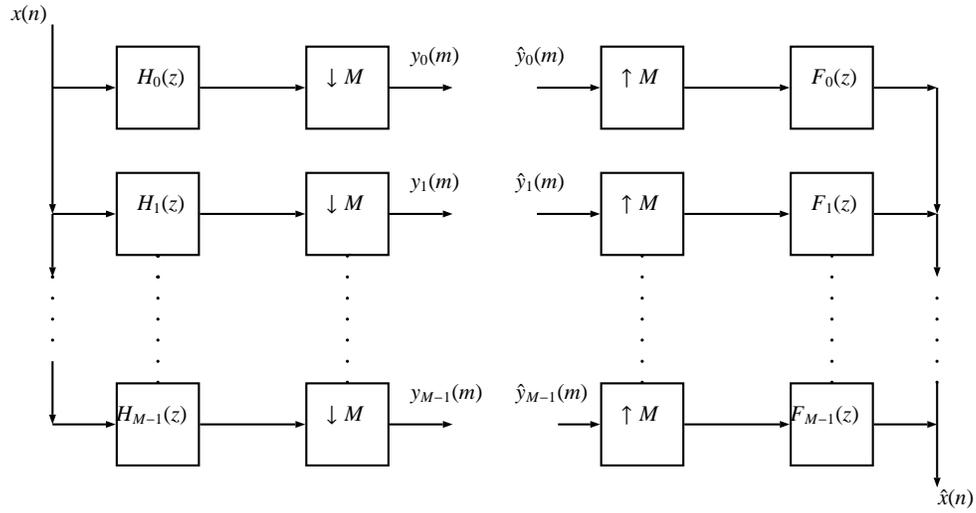

The analysis filters are represented as \citep{fliege1994multirate}
\begin {equation}
H_{k}(z) = H(zW_{M}^{k}) \\ , \quad k = 0,1,......M-1
\end {equation}

The synthesis filters are defined as \citep{fliege1994multirate}

\begin {equation}
F_{k}(z) = MH(zW_{M}^{k}) 
\end {equation}

where
\begin {equation}
 W_{M} = exp[-j(\frac{2\pi}{M})]
\end {equation}
  
The reconstructed signal at the output of the FB is defined as \citep{fliege1994multirate}

\begin {eqnarray}
\hat{X}(z) = \frac{1}{M} \sum_{k=0}^{M - 1}F_{k}(z)  \sum_{l=0}^{M-1} H(zW_{M}^{k+1})X(zW_{M}^{l})
\end {eqnarray}

assuming $X(z)$ is the input signal.

Since the prototype filter is band-limited to $2\pi/M$, all the non-adjacent alias components are removed. If we consider only adjacent alias components, the reconstructed signal can be written as \citep{fliege1994multirate}

 \begin {eqnarray}
\hat{X}(z) = \frac{1}{M} \sum_{k=0}^{M - 1}F_{k}(z)  \sum_{l=-1}^{1} H(zW_{M}^{k+1})X(zW_{M}^{l})
\end {eqnarray}

From equation (4), we can understand that the reconstructed signal contains alias components.
There is no way in the structure of the DFT FB to compensate for the alias signals.
To overcome this disadvantage some modifications have been done in the DFT FB, which result in the MDFT FB \citep{fliege1994modified}.

\par The MDFT FB can be derived from a complex modulated FB by decimating the sampling rate with and without a delay of $M/2$ samples and using either the real or the imaginary part, alternately, in the sub-bands as shown in Fig.2. \citep{fliege1994modified}. These modifications eliminate directly adjacent alias spectra which are the main components of aliasing in the DFT FB. Non-adjacent alias terms can be made small by selecting high stopband attenuation in the design of the prototype filter. This will give NPR in MDFT FB.

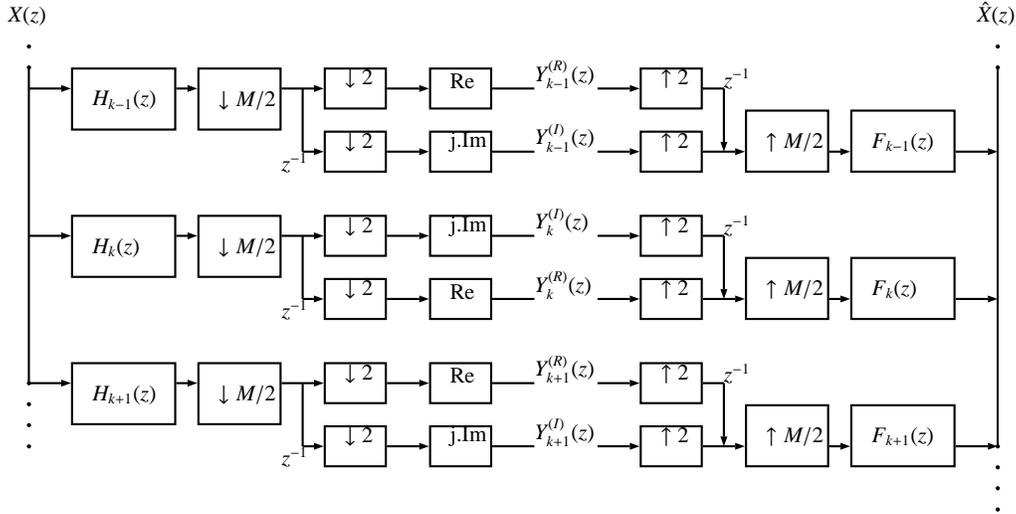
\begin{figure}[here]
\centering

\psscalebox{0.7 0.7} 
{
\begin{pspicture}(0,-4.7947116)(19.62,4.7947116)
\psframe[linecolor=black, linewidth=0.04, dimen=outer](5.2,3.6447115)(3.6,2.4447117)
\psframe[linecolor=black, linewidth=0.04, dimen=outer](7.2,3.6447115)(6.0,2.8447115)
\psframe[linecolor=black, linewidth=0.04, dimen=outer](7.2,2.4447117)(6.0,1.6447116)
\psframe[linecolor=black, linewidth=0.04, dimen=outer](9.2,3.6447115)(8.0,2.8447115)
\psframe[linecolor=black, linewidth=0.04, dimen=outer](9.2,2.4447117)(8.0,1.6447116)
\psframe[linecolor=black, linewidth=0.04, dimen=outer](13.2,3.6447115)(12.0,2.8447115)
\psframe[linecolor=black, linewidth=0.04, dimen=outer](13.2,2.4447117)(12.0,1.6447116)
\psframe[linecolor=black, linewidth=0.04, dimen=outer](15.6,2.8447115)(14.0,1.6447116)
\psframe[linecolor=black, linewidth=0.04, dimen=outer](5.2,0.8447116)(3.6,-0.3552884)
\psframe[linecolor=black, linewidth=0.04, dimen=outer](7.2,0.8447116)(6.0,0.04471161)
\psframe[linecolor=black, linewidth=0.04, dimen=outer](7.2,-0.3552884)(6.0,-1.1552883)
\psframe[linecolor=black, linewidth=0.04, dimen=outer](9.2,0.8447116)(8.0,0.04471161)
\psframe[linecolor=black, linewidth=0.04, dimen=outer](9.2,-0.3552884)(8.0,-1.1552883)
\psframe[linecolor=black, linewidth=0.04, dimen=outer](13.2,0.8447116)(12.0,0.04471161)
\psframe[linecolor=black, linewidth=0.04, dimen=outer](13.2,-0.3552884)(12.0,-1.1552883)
\psframe[linecolor=black, linewidth=0.04, dimen=outer](5.2,-1.9552884)(3.6,-3.1552885)
\psframe[linecolor=black, linewidth=0.04, dimen=outer](7.2,-1.9552884)(6.0,-2.7552884)
\psframe[linecolor=black, linewidth=0.04, dimen=outer](7.2,-3.1552885)(6.0,-3.9552884)
\psframe[linecolor=black, linewidth=0.04, dimen=outer](9.2,-1.9552884)(8.0,-2.7552884)
\psframe[linecolor=black, linewidth=0.04, dimen=outer](9.2,-3.1552885)(8.0,-3.9552884)
\psframe[linecolor=black, linewidth=0.04, dimen=outer](13.2,-1.9552884)(12.0,-2.7552884)
\psframe[linecolor=black, linewidth=0.04, dimen=outer](13.2,-3.1552885)(12.0,-3.9552884)
\psframe[linecolor=black, linewidth=0.04, dimen=outer](15.6,-2.7552884)(14.0,-3.9552884)
\rput[bl](4.0,2.8447115){$\downarrow M/2$}
\rput[bl](6.4,3.2447116){$\downarrow 2$}
\psline[linecolor=black, linewidth=0.04](0.4,3.6447115)(0.4,-2.3552885)
\psline[linecolor=black, linewidth=0.04, arrowsize=0.05291666666666668cm 2.0,arrowlength=1.4,arrowinset=0.0]{->}(0.4,-2.3552885)(1.2,-2.3552885)
\psline[linecolor=black, linewidth=0.04, arrowsize=0.05291666666666668cm 2.0,arrowlength=1.4,arrowinset=0.0]{->}(0.4,0.4447116)(1.2,0.4447116)
\psline[linecolor=black, linewidth=0.04, arrowsize=0.05291666666666668cm 2.0,arrowlength=1.4,arrowinset=0.0]{->}(0.4,3.2447116)(1.2,3.2447116)
\psline[linecolor=black, linewidth=0.04, arrowsize=0.05291666666666668cm 2.0,arrowlength=1.4,arrowinset=0.0]{->}(5.2,3.2447116)(6.0,3.2447116)
\psline[linecolor=black, linewidth=0.04, arrowsize=0.05291666666666668cm 2.0,arrowlength=1.4,arrowinset=0.0]{->}(5.2,0.4447116)(6.0,0.4447116)
\psline[linecolor=black, linewidth=0.04, arrowsize=0.05291666666666668cm 2.0,arrowlength=1.4,arrowinset=0.0]{->}(5.2,-2.3552885)(6.0,-2.3552885)
\psline[linecolor=black, linewidth=0.04, arrowsize=0.05291666666666668cm 2.0,arrowlength=1.4,arrowinset=0.0]{->}(7.2,3.2447116)(8.0,3.2447116)
\psline[linecolor=black, linewidth=0.04, arrowsize=0.05291666666666668cm 2.0,arrowlength=1.4,arrowinset=0.0]{->}(7.2,2.0447116)(8.0,2.0447116)
\psline[linecolor=black, linewidth=0.04, arrowsize=0.05291666666666668cm 2.0,arrowlength=1.4,arrowinset=0.0]{->}(7.2,0.4447116)(8.0,0.4447116)
\psline[linecolor=black, linewidth=0.04, arrowsize=0.05291666666666668cm 2.0,arrowlength=1.4,arrowinset=0.0]{->}(7.2,-0.75528836)(8.0,-0.75528836)
\psline[linecolor=black, linewidth=0.04, arrowsize=0.05291666666666668cm 2.0,arrowlength=1.4,arrowinset=0.0]{->}(7.2,-2.3552885)(8.0,-2.3552885)
\psline[linecolor=black, linewidth=0.04, arrowsize=0.05291666666666668cm 2.0,arrowlength=1.4,arrowinset=0.0]{->}(7.2,-3.5552883)(8.0,-3.5552883)
\psline[linecolor=black, linewidth=0.04, arrowsize=0.05291666666666668cm 2.0,arrowlength=1.4,arrowinset=0.0]{->}(13.2,2.0447116)(14.0,2.0447116)
\psline[linecolor=black, linewidth=0.04, arrowsize=0.05291666666666668cm 2.0,arrowlength=1.4,arrowinset=0.0]{->}(13.2,-0.75528836)(14.0,-0.75528836)
\psline[linecolor=black, linewidth=0.04, arrowsize=0.05291666666666668cm 2.0,arrowlength=1.4,arrowinset=0.0]{->}(13.2,-3.5552883)(14.0,-3.5552883)
\psline[linecolor=black, linewidth=0.04](18.8,3.6447115)(18.8,-3.5552883)
\psline[linecolor=black, linewidth=0.04, arrowsize=0.05291666666666668cm 2.0,arrowlength=1.4,arrowinset=0.0]{->}(18.0,2.0447116)(18.8,2.0447116)
\psline[linecolor=black, linewidth=0.04, arrowsize=0.05291666666666668cm 2.0,arrowlength=1.4,arrowinset=0.0]{->}(18.0,-0.75528836)(18.8,-0.75528836)
\psline[linecolor=black, linewidth=0.04, arrowsize=0.05291666666666668cm 2.0,arrowlength=1.4,arrowinset=0.0]{->}(18.0,-3.5552883)(18.8,-3.5552883)
\psdots[linecolor=black, dotsize=0.08](0.4,3.6447115)
\psdots[linecolor=black, dotsize=0.08](0.4,4.0447116)
\psdots[linecolor=black, dotsize=0.08](0.4,-2.7552884)
\psdots[linecolor=black, dotsize=0.08](0.4,-3.1552885)
\psdots[linecolor=black, dotsize=0.08](0.4,-3.5552883)
\psdots[linecolor=black, dotsize=0.08](18.8,-3.5552883)
\psdots[linecolor=black, dotsize=0.08](18.8,-3.9552884)
\psdots[linecolor=black, dotsize=0.08](18.8,-4.3552885)
\psdots[linecolor=black, dotsize=0.08](18.8,4.0447116)
\psdots[linecolor=black, dotsize=0.08](18.8,-4.7552886)
\rput[bl](8.4,3.2447116){Re}
\rput[bl](6.4,2.0447116){$\downarrow 2$}
\rput[bl](8.4,2.0447116){j.Im}
\rput[bl](1.6,0.04471161){$H_{k}(z)$}
\psframe[linecolor=black, linewidth=0.04, dimen=outer](3.2,3.6447115)(1.2,2.4447117)
\rput[bl](1.6,2.8447115){$H_{k-1}(z)$}
\psframe[linecolor=black, linewidth=0.04, dimen=outer](3.2,0.8447116)(1.2,-0.3552884)
\psframe[linecolor=black, linewidth=0.04, dimen=outer](3.2,-1.9552884)(1.2,-3.1552885)
\psframe[linecolor=black, linewidth=0.04, dimen=outer](18.0,2.8447115)(16.0,1.6447116)
\psframe[linecolor=black, linewidth=0.04, dimen=outer](18.0,0.04471161)(16.0,-1.1552883)
\psframe[linecolor=black, linewidth=0.04, dimen=outer](18.0,-2.7552884)(16.0,-3.9552884)
\rput[bl](1.6,-2.7552884){$H_{k+1}(z)$}
\rput[bl](4.0,0.04471161){$\downarrow M/2$}
\rput[bl](4.0,-2.7552884){$\downarrow M/2$}
\rput[bl](6.4,-0.75528836){$\downarrow 2$}
\rput[bl](6.4,-2.3552885){$\downarrow 2$}
\rput[bl](6.4,-3.5552883){$\downarrow 2$}
\rput[bl](6.4,0.4447116){$\downarrow 2$}
\rput[bl](8.4,0.4447116){j.Im}
\rput[bl](8.4,-0.75528836){Re}
\rput[bl](8.4,-2.3552885){Re}
\rput[bl](8.4,-3.5552883){j.Im}
\rput[bl](12.4,3.2447116){$\uparrow 2$}
\rput[bl](12.4,2.0447116){$\uparrow 2$}
\rput[bl](12.4,0.4447116){$\uparrow 2$}
\rput[bl](12.4,-0.75528836){$\uparrow 2$}
\rput[bl](12.4,-2.3552885){$\uparrow 2$}
\rput[bl](12.4,-3.5552883){$\uparrow 2$}
\rput[bl](14.4,2.0447116){$\uparrow M/2$}
\rput[bl](14.4,-0.75528836){$\uparrow M/2$}
\rput[bl](14.4,-3.5552883){$\uparrow M/2$}
\rput[bl](16.4,2.0447116){$F_{k-1}(z)$}
\rput[bl](16.4,-0.75528836){$F_{k}(z)$}
\rput[bl](16.4,-3.5552883){$F_{k+1}(z)$}
\psline[linecolor=black, linewidth=0.04](5.6,3.2447116)(5.6,2.0447116)
\psline[linecolor=black, linewidth=0.04](5.6,0.4447116)(5.6,-0.75528836)
\psline[linecolor=black, linewidth=0.04](5.6,-2.3552885)(5.6,-3.5552883)
\psline[linecolor=black, linewidth=0.04](13.2,3.2447116)(13.2,3.2447116)
\psline[linecolor=black, linewidth=0.04](13.2,3.2447116)(13.2,3.2447116)(13.6,3.2447116)
\psline[linecolor=black, linewidth=0.04](13.2,0.4447116)(13.6,0.4447116)
\psline[linecolor=black, linewidth=0.04](13.2,-2.3552885)(13.6,-2.3552885)
\psline[linecolor=black, linewidth=0.04, arrowsize=0.05291666666666668cm 2.0,arrowlength=1.4,arrowinset=0.0]{->}(3.2,3.2447116)(3.6,3.2447116)
\psline[linecolor=black, linewidth=0.04, arrowsize=0.05291666666666668cm 2.0,arrowlength=1.4,arrowinset=0.0]{->}(3.2,0.4447116)(3.6,0.4447116)
\psline[linecolor=black, linewidth=0.04, arrowsize=0.05291666666666668cm 2.0,arrowlength=1.4,arrowinset=0.0]{->}(3.2,-2.3552885)(3.6,-2.3552885)
\psframe[linecolor=black, linewidth=0.04, dimen=outer](15.6,0.04471161)(14.0,-1.1552883)
\psline[linecolor=black, linewidth=0.04, arrowsize=0.05291666666666668cm 2.0,arrowlength=1.4,arrowinset=0.0]{->}(15.6,-0.75528836)(16.0,-0.75528836)
\psline[linecolor=black, linewidth=0.04, arrowsize=0.05291666666666668cm 2.0,arrowlength=1.4,arrowinset=0.0]{->}(15.6,-3.5552883)(16.0,-3.5552883)
\psline[linecolor=black, linewidth=0.04, arrowsize=0.05291666666666668cm 2.0,arrowlength=1.4,arrowinset=0.0]{->}(15.6,2.0447116)(16.0,2.0447116)
\psline[linecolor=black, linewidth=0.04, arrowsize=0.05291666666666668cm 2.0,arrowlength=1.4,arrowinset=0.0]{->}(5.6,2.0447116)(6.0,2.0447116)
\psline[linecolor=black, linewidth=0.04, arrowsize=0.05291666666666668cm 2.0,arrowlength=1.4,arrowinset=0.0]{->}(5.6,-0.75528836)(6.0,-0.75528836)
\psline[linecolor=black, linewidth=0.04, arrowsize=0.05291666666666668cm 2.0,arrowlength=1.4,arrowinset=0.0]{->}(5.6,-3.5552883)(6.0,-3.5552883)
\psline[linecolor=black, linewidth=0.04, arrowsize=0.05291666666666668cm 2.0,arrowlength=1.4,arrowinset=0.0]{->}(13.6,3.2447116)(13.6,2.0447116)
\psline[linecolor=black, linewidth=0.04, arrowsize=0.05291666666666668cm 2.0,arrowlength=1.4,arrowinset=0.0]{->}(13.6,0.4447116)(13.6,-0.75528836)
\psline[linecolor=black, linewidth=0.04, arrowsize=0.05291666666666668cm 2.0,arrowlength=1.4,arrowinset=0.0]{->}(13.6,-2.3552885)(13.6,-3.5552883)
\rput[bl](10.0,3.2447116){$Y_{k-1}^{(R)}(z)$}
\rput[bl](10.0,2.0447116){$Y_{k-1}^{(I)}(z)$}
\rput[bl](10.0,0.4447116){$Y_{k}^{(I)}(z)$}
\rput[bl](10.0,-0.75528836){$Y_{k}^{(R)}(z)$}
\rput[bl](10.0,-2.3552885){$Y_{k+1}^{(R)}(z)$}
\rput[bl](10.0,-3.5552883){$Y_{k+1}^{(I)}(z)$}
\rput[bl](13.6,3.2447116){$z^{-1}$}
\rput[bl](13.6,0.4447116){$z^{-1}$}
\rput[bl](13.6,-2.3552885){$z^{-1}$}
\rput[bl](5.2,1.6447116){$z^{-1}$}
\rput[bl](5.2,-1.1552883){$z^{-1}$}
\rput[bl](5.2,-3.9552884){$z^{-1}$}
\rput[bl](0.0,4.4447117){$X(z)$}
\rput[bl](18.4,4.4447117){$\hat{X}(z)$}
\psdots[linecolor=black, dotsize=0.08](18.8,3.6447115)
\psdots[linecolor=black, dotsize=0.08](0.4,-2.3552885)
\psline[linecolor=black, linewidth=0.04](9.2,3.2447116)(10.0,3.2447116)
\psline[linecolor=black, linewidth=0.04](9.2,2.0447116)(10.0,2.0447116)
\psline[linecolor=black, linewidth=0.04](9.2,0.4447116)(10.0,0.4447116)
\psline[linecolor=black, linewidth=0.04](9.2,-0.75528836)(10.0,-0.75528836)
\psline[linecolor=black, linewidth=0.04](9.2,-2.3552885)(10.0,-2.3552885)
\psline[linecolor=black, linewidth=0.04](9.2,-3.5552883)(10.0,-3.5552883)
\psline[linecolor=black, linewidth=0.04, arrowsize=0.05291666666666667cm 2.0,arrowlength=1.4,arrowinset=0.0]{->}(11.2,0.4447116)(12.0,0.4447116)
\psline[linecolor=black, linewidth=0.04, arrowsize=0.05291666666666667cm 2.0,arrowlength=1.4,arrowinset=0.0]{->}(11.2,-0.75528836)(12.0,-0.75528836)
\psline[linecolor=black, linewidth=0.04, arrowsize=0.05291666666666667cm 2.0,arrowlength=1.4,arrowinset=0.0]{->}(11.2,-3.5552883)(12.0,-3.5552883)
\psline[linecolor=black, linewidth=0.04, arrowsize=0.05291666666666667cm 2.0,arrowlength=1.4,arrowinset=0.0]{->}(11.2,3.2447116)(12.0,3.2447116)
\psline[linecolor=black, linewidth=0.04, arrowsize=0.05291666666666667cm 2.0,arrowlength=1.4,arrowinset=0.0]{->}(11.2,2.0447116)(12.0,2.0447116)
\psline[linecolor=black, linewidth=0.04, arrowsize=0.05291666666666667cm 2.0,arrowlength=1.4,arrowinset=0.0]{->}(11.2,-2.3552885)(12.0,-2.3552885)
\end{pspicture}
}

\caption{Modified DFT filter bank}
\label{mdft}

\end{figure}

 \par The output signal $\hat{X} (z)$ in the MDFT FB is given as \citep{sakthivel2014design,karp1999modified}

\begin {eqnarray}
\label{eq3}
\hat{X} (z) = \frac{z^{-M/2}}{M} \sum_{k=0}^{M - 1} \sum_{l=0}^{M/2-1}F_{k}(z) H_{k}(zW_{M}^{2l})X(zW_{M}^{2l})
\end {eqnarray}

\par If the prototype filter is designed as a linear phase filter, all the analysis and synthesis filters of MDFT FB will have linear phase. Hence, the MDFT FB is free from phase distortions. The power complementary property of adjacent channels and the stop band attenuation depend on the prototype filter design. In the MDFT FB, the amplitude distortion function is given as \citep{fliege1994multirate,pp} 

\begin {equation}
T_{dist}(z) = \frac{1}{M}\sum_{k=0}^{M - 1}F_{k}(z)H_{k}(z) 
\end {equation}

All the analysis and synthesis filters of the MDFT FB are obtained from the same prototype filter utilizing complex modulation. Hence the problem of designing the MDFT FB decreases to the problem of designing a single prototype filter. 

\label{sec:2} 
\section{Proposed Design of MDFT non-uniform FB }

The non-uniform FB decompose the input signal into subbands of unequal bandwidths. The structure of an $\tilde{M}$ channel non-uniform MDFT FB is shown in Fig.3. A set of $\tilde{M}$ analysis filters $\tilde{H}_{k}(z)$, $0 \leq k \leq \tilde{M}-1$ decomposes the input signal into $\tilde{M}$ subbands. A set of synthesis filters $\tilde{F}_{k}(z), 0\leq k \leq \tilde{M}-1$ combines the $\tilde{M}$ subband signals. The decimation ratios are not equal in all the subbands. The $\tilde{M}$ channel non-uniform design is obtained from the $M$-channel uniform MDFT FB by merging adjacent channels. For maximally decimated FB, the decimation factors should satisfy the condition $\sum_{k=0}^{\tilde{M}-1}\frac{1}{M_{k}}=1$.

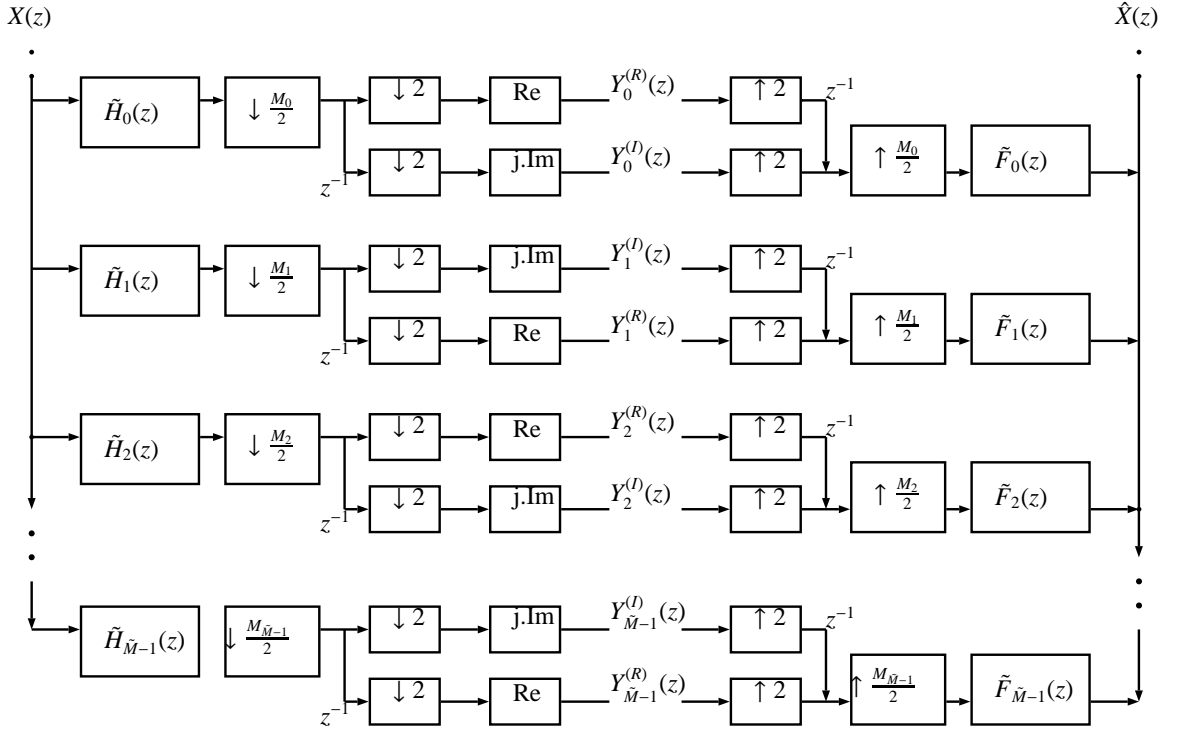
\begin{figure}[here]
\centering

\psscalebox{0.8 0.8} 
{
\begin{pspicture}(0,-5.975)(19.62,5.975)
\psframe[linecolor=black, linewidth=0.04, dimen=outer](5.2,4.825)(3.6,3.625)
\psframe[linecolor=black, linewidth=0.04, dimen=outer](7.2,4.825)(6.0,4.025)
\psframe[linecolor=black, linewidth=0.04, dimen=outer](7.2,3.625)(6.0,2.825)
\psframe[linecolor=black, linewidth=0.04, dimen=outer](9.2,4.825)(8.0,4.025)
\psframe[linecolor=black, linewidth=0.04, dimen=outer](9.2,3.625)(8.0,2.825)
\psframe[linecolor=black, linewidth=0.04, dimen=outer](13.2,4.825)(12.0,4.025)
\psframe[linecolor=black, linewidth=0.04, dimen=outer](13.2,3.625)(12.0,2.825)
\psframe[linecolor=black, linewidth=0.04, dimen=outer](15.6,4.025)(14.0,2.825)
\psframe[linecolor=black, linewidth=0.04, dimen=outer](5.2,2.025)(3.6,0.825)
\psframe[linecolor=black, linewidth=0.04, dimen=outer](7.2,2.025)(6.0,1.225)
\psframe[linecolor=black, linewidth=0.04, dimen=outer](7.2,0.825)(6.0,0.025)
\psframe[linecolor=black, linewidth=0.04, dimen=outer](9.2,2.025)(8.0,1.225)
\psframe[linecolor=black, linewidth=0.04, dimen=outer](9.2,0.825)(8.0,0.025)
\psframe[linecolor=black, linewidth=0.04, dimen=outer](13.2,2.025)(12.0,1.225)
\psframe[linecolor=black, linewidth=0.04, dimen=outer](13.2,0.825)(12.0,0.025)
\psframe[linecolor=black, linewidth=0.04, dimen=outer](5.2,-0.775)(3.6,-1.975)
\psframe[linecolor=black, linewidth=0.04, dimen=outer](7.2,-0.775)(6.0,-1.575)
\psframe[linecolor=black, linewidth=0.04, dimen=outer](7.2,-1.975)(6.0,-2.775)
\psframe[linecolor=black, linewidth=0.04, dimen=outer](9.2,-0.775)(8.0,-1.575)
\psframe[linecolor=black, linewidth=0.04, dimen=outer](9.2,-1.975)(8.0,-2.775)
\psframe[linecolor=black, linewidth=0.04, dimen=outer](13.2,-0.775)(12.0,-1.575)
\psframe[linecolor=black, linewidth=0.04, dimen=outer](13.2,-1.975)(12.0,-2.775)
\psframe[linecolor=black, linewidth=0.04, dimen=outer](15.6,-1.575)(14.0,-2.775)
\rput[bl](6.4,4.425){$\downarrow 2$}
\psline[linecolor=black, linewidth=0.04](0.4,4.825)(0.4,-1.175)
\psline[linecolor=black, linewidth=0.04, arrowsize=0.05291666666666668cm 2.0,arrowlength=1.4,arrowinset=0.0]{->}(0.4,-1.175)(1.2,-1.175)
\psline[linecolor=black, linewidth=0.04, arrowsize=0.05291666666666668cm 2.0,arrowlength=1.4,arrowinset=0.0]{->}(0.4,1.625)(1.2,1.625)
\psline[linecolor=black, linewidth=0.04, arrowsize=0.05291666666666668cm 2.0,arrowlength=1.4,arrowinset=0.0]{->}(0.4,4.425)(1.2,4.425)
\psline[linecolor=black, linewidth=0.04, arrowsize=0.05291666666666668cm 2.0,arrowlength=1.4,arrowinset=0.0]{->}(5.2,4.425)(6.0,4.425)
\psline[linecolor=black, linewidth=0.04, arrowsize=0.05291666666666668cm 2.0,arrowlength=1.4,arrowinset=0.0]{->}(5.2,1.625)(6.0,1.625)
\psline[linecolor=black, linewidth=0.04, arrowsize=0.05291666666666668cm 2.0,arrowlength=1.4,arrowinset=0.0]{->}(5.2,-1.175)(6.0,-1.175)
\psline[linecolor=black, linewidth=0.04, arrowsize=0.05291666666666668cm 2.0,arrowlength=1.4,arrowinset=0.0]{->}(7.2,4.425)(8.0,4.425)
\psline[linecolor=black, linewidth=0.04, arrowsize=0.05291666666666668cm 2.0,arrowlength=1.4,arrowinset=0.0]{->}(7.2,3.225)(8.0,3.225)
\psline[linecolor=black, linewidth=0.04, arrowsize=0.05291666666666668cm 2.0,arrowlength=1.4,arrowinset=0.0]{->}(7.2,1.625)(8.0,1.625)
\psline[linecolor=black, linewidth=0.04, arrowsize=0.05291666666666668cm 2.0,arrowlength=1.4,arrowinset=0.0]{->}(7.2,0.425)(8.0,0.425)
\psline[linecolor=black, linewidth=0.04, arrowsize=0.05291666666666668cm 2.0,arrowlength=1.4,arrowinset=0.0]{->}(7.2,-1.175)(8.0,-1.175)
\psline[linecolor=black, linewidth=0.04, arrowsize=0.05291666666666668cm 2.0,arrowlength=1.4,arrowinset=0.0]{->}(7.2,-2.375)(8.0,-2.375)
\psline[linecolor=black, linewidth=0.04, arrowsize=0.05291666666666668cm 2.0,arrowlength=1.4,arrowinset=0.0]{->}(13.2,3.225)(14.0,3.225)
\psline[linecolor=black, linewidth=0.04, arrowsize=0.05291666666666668cm 2.0,arrowlength=1.4,arrowinset=0.0]{->}(13.2,0.425)(14.0,0.425)
\psline[linecolor=black, linewidth=0.04, arrowsize=0.05291666666666668cm 2.0,arrowlength=1.4,arrowinset=0.0]{->}(13.2,-2.375)(14.0,-2.375)
\psline[linecolor=black, linewidth=0.04](18.8,4.825)(18.8,-2.375)
\psline[linecolor=black, linewidth=0.04, arrowsize=0.05291666666666668cm 2.0,arrowlength=1.4,arrowinset=0.0]{->}(18.0,3.225)(18.8,3.225)
\psline[linecolor=black, linewidth=0.04, arrowsize=0.05291666666666668cm 2.0,arrowlength=1.4,arrowinset=0.0]{->}(18.0,0.425)(18.8,0.425)
\psline[linecolor=black, linewidth=0.04, arrowsize=0.05291666666666668cm 2.0,arrowlength=1.4,arrowinset=0.0]{->}(18.0,-2.375)(18.8,-2.375)
\psdots[linecolor=black, dotsize=0.08](0.4,4.825)
\psdots[linecolor=black, dotsize=0.08](0.4,5.225)
\psdots[linecolor=black, dotsize=0.08](18.8,-2.375)
\psdots[linecolor=black, dotsize=0.08](18.8,5.225)
\rput[bl](8.4,4.425){Re}
\rput[bl](6.4,3.225){$\downarrow 2$}
\rput[bl](8.4,3.225){j.Im}
\psframe[linecolor=black, linewidth=0.04, dimen=outer](3.2,-0.775)(1.2,-1.975)
\psframe[linecolor=black, linewidth=0.04, dimen=outer](18.0,4.025)(16.0,2.825)
\psframe[linecolor=black, linewidth=0.04, dimen=outer](18.0,1.225)(16.0,0.025)
\psframe[linecolor=black, linewidth=0.04, dimen=outer](18.0,-1.575)(16.0,-2.775)
\rput[bl](6.4,0.425){$\downarrow 2$}
\rput[bl](6.4,-1.175){$\downarrow 2$}
\rput[bl](6.4,-2.375){$\downarrow 2$}
\rput[bl](6.4,1.625){$\downarrow 2$}
\rput[bl](8.4,1.625){j.Im}
\rput[bl](8.4,0.425){Re}
\rput[bl](8.4,-1.175){Re}
\rput[bl](8.4,-2.375){j.Im}
\rput[bl](12.4,4.425){$\uparrow 2$}
\rput[bl](12.4,3.225){$\uparrow 2$}
\rput[bl](12.4,1.625){$\uparrow 2$}
\rput[bl](12.4,0.425){$\uparrow 2$}
\rput[bl](12.4,-1.175){$\uparrow 2$}
\rput[bl](12.4,-2.375){$\uparrow 2$}
\psline[linecolor=black, linewidth=0.04](5.6,4.425)(5.6,3.225)
\psline[linecolor=black, linewidth=0.04](5.6,1.625)(5.6,0.425)
\psline[linecolor=black, linewidth=0.04](5.6,-1.175)(5.6,-2.375)
\psline[linecolor=black, linewidth=0.04](13.2,4.425)(13.2,4.425)
\psline[linecolor=black, linewidth=0.04](13.2,4.425)(13.2,4.425)(13.6,4.425)
\psline[linecolor=black, linewidth=0.04](13.2,1.625)(13.6,1.625)
\psline[linecolor=black, linewidth=0.04](13.2,-1.175)(13.6,-1.175)
\psline[linecolor=black, linewidth=0.04, arrowsize=0.05291666666666668cm 2.0,arrowlength=1.4,arrowinset=0.0]{->}(3.2,4.425)(3.6,4.425)
\psline[linecolor=black, linewidth=0.04, arrowsize=0.05291666666666668cm 2.0,arrowlength=1.4,arrowinset=0.0]{->}(3.2,1.625)(3.6,1.625)
\psline[linecolor=black, linewidth=0.04, arrowsize=0.05291666666666668cm 2.0,arrowlength=1.4,arrowinset=0.0]{->}(3.2,-1.175)(3.6,-1.175)
\psframe[linecolor=black, linewidth=0.04, dimen=outer](15.6,1.225)(14.0,0.025)
\psline[linecolor=black, linewidth=0.04, arrowsize=0.05291666666666668cm 2.0,arrowlength=1.4,arrowinset=0.0]{->}(15.6,0.425)(16.0,0.425)
\psline[linecolor=black, linewidth=0.04, arrowsize=0.05291666666666668cm 2.0,arrowlength=1.4,arrowinset=0.0]{->}(15.6,-2.375)(16.0,-2.375)
\psline[linecolor=black, linewidth=0.04, arrowsize=0.05291666666666668cm 2.0,arrowlength=1.4,arrowinset=0.0]{->}(15.6,3.225)(16.0,3.225)
\psline[linecolor=black, linewidth=0.04, arrowsize=0.05291666666666668cm 2.0,arrowlength=1.4,arrowinset=0.0]{->}(5.6,3.225)(6.0,3.225)
\psline[linecolor=black, linewidth=0.04, arrowsize=0.05291666666666668cm 2.0,arrowlength=1.4,arrowinset=0.0]{->}(5.6,0.425)(6.0,0.425)
\psline[linecolor=black, linewidth=0.04, arrowsize=0.05291666666666668cm 2.0,arrowlength=1.4,arrowinset=0.0]{->}(5.6,-2.375)(6.0,-2.375)
\psline[linecolor=black, linewidth=0.04, arrowsize=0.05291666666666668cm 2.0,arrowlength=1.4,arrowinset=0.0]{->}(13.6,4.425)(13.6,3.225)
\psline[linecolor=black, linewidth=0.04, arrowsize=0.05291666666666668cm 2.0,arrowlength=1.4,arrowinset=0.0]{->}(13.6,1.625)(13.6,0.425)
\psline[linecolor=black, linewidth=0.04, arrowsize=0.05291666666666668cm 2.0,arrowlength=1.4,arrowinset=0.0]{->}(13.6,-1.175)(13.6,-2.375)
\rput[bl](10.0,4.425){$Y_{0}^{(R)}(z)$}
\rput[bl](10.0,3.225){$Y_{0}^{(I)}(z)$}
\rput[bl](10.0,1.625){$Y_{1}^{(I)}(z)$}
\rput[bl](10.0,0.425){$Y_{1}^{(R)}(z)$}
\rput[bl](10.0,-1.175){$Y_{2}^{(R)}(z)$}
\rput[bl](10.0,-2.375){$Y_{2}^{(I)}(z)$}
\rput[bl](13.6,4.425){$z^{-1}$}
\rput[bl](13.6,1.625){$z^{-1}$}
\rput[bl](13.6,-1.175){$z^{-1}$}
\rput[bl](5.2,2.825){$z^{-1}$}
\rput[bl](5.2,0.025){$z^{-1}$}
\rput[bl](5.2,-2.775){$z^{-1}$}
\rput[bl](0.0,5.625){$X(z)$}
\rput[bl](18.4,5.625){$\hat{X}(z)$}
\psdots[linecolor=black, dotsize=0.08](18.8,4.825)
\psdots[linecolor=black, dotsize=0.08](0.4,-1.175)
\psline[linecolor=black, linewidth=0.04](9.2,4.425)(10.0,4.425)
\psline[linecolor=black, linewidth=0.04](9.2,3.225)(10.0,3.225)
\psline[linecolor=black, linewidth=0.04](9.2,1.625)(10.0,1.625)
\psline[linecolor=black, linewidth=0.04](9.2,0.425)(10.0,0.425)
\psline[linecolor=black, linewidth=0.04](9.2,-1.175)(10.0,-1.175)
\psline[linecolor=black, linewidth=0.04](9.2,-2.375)(10.0,-2.375)
\psline[linecolor=black, linewidth=0.04, arrowsize=0.05291666666666668cm 2.0,arrowlength=1.4,arrowinset=0.0]{->}(11.2,1.625)(12.0,1.625)
\psline[linecolor=black, linewidth=0.04, arrowsize=0.05291666666666668cm 2.0,arrowlength=1.4,arrowinset=0.0]{->}(11.2,0.425)(12.0,0.425)
\psline[linecolor=black, linewidth=0.04, arrowsize=0.05291666666666668cm 2.0,arrowlength=1.4,arrowinset=0.0]{->}(11.2,-2.375)(12.0,-2.375)
\psline[linecolor=black, linewidth=0.04, arrowsize=0.05291666666666668cm 2.0,arrowlength=1.4,arrowinset=0.0]{->}(11.2,4.425)(12.0,4.425)
\psline[linecolor=black, linewidth=0.04, arrowsize=0.05291666666666668cm 2.0,arrowlength=1.4,arrowinset=0.0]{->}(11.2,3.225)(12.0,3.225)
\psline[linecolor=black, linewidth=0.04, arrowsize=0.05291666666666668cm 2.0,arrowlength=1.4,arrowinset=0.0]{->}(11.2,-1.175)(12.0,-1.175)
\psframe[linecolor=black, linewidth=0.04, dimen=outer](3.2,4.825)(1.2,3.625)
\rput[bl](1.6,4.025){$\tilde{H}_{0}(z)$}
\psframe[linecolor=black, linewidth=0.04, dimen=outer](3.2,2.025)(1.2,0.825)
\rput[bl](1.6,1.225){$\tilde{H}_{1}(z)$}
\rput[bl](1.6,-1.575){$\tilde{H}_{2}(z)$}
\rput[bl](4.0,4.025){$\downarrow \frac{M_{0}}{2}$}
\rput[bl](4.0,1.225){$\downarrow \frac{M_{1}}{2}$}
\rput[bl](4.0,-1.575){$\downarrow \frac{M_{2}}{2}$}
\rput[bl](16.4,3.225){$\tilde{F}_{0}(z)$}
\rput[bl](16.4,0.425){$\tilde{F}_{1}(z)$}
\rput[bl](16.4,-2.375){$\tilde{F}_{2}(z)$}
\rput[bl](14.4,3.225){$\uparrow \frac{M_{0}}{2}$}
\rput[bl](14.4,0.425){$\uparrow \frac{M_{1}}{2}$}
\rput[bl](14.4,-2.375){$\uparrow \frac{M_{2}}{2}$}
\psframe[linecolor=black, linewidth=0.04, dimen=outer](3.2,-3.975)(1.2,-5.175)
\psframe[linecolor=black, linewidth=0.04, dimen=outer](5.2,-3.975)(3.6,-5.175)
\psframe[linecolor=black, linewidth=0.04, dimen=outer](7.2,-3.975)(6.0,-4.775)
\psframe[linecolor=black, linewidth=0.04, dimen=outer](7.2,-5.175)(6.0,-5.975)
\psframe[linecolor=black, linewidth=0.04, dimen=outer](9.2,-3.975)(8.0,-4.775)
\psframe[linecolor=black, linewidth=0.04, dimen=outer](9.2,-5.175)(8.0,-5.975)
\psframe[linecolor=black, linewidth=0.04, dimen=outer](13.2,-3.975)(12.0,-4.775)
\psframe[linecolor=black, linewidth=0.04, dimen=outer](13.2,-5.175)(12.0,-5.975)
\psframe[linecolor=black, linewidth=0.04, dimen=outer](15.6,-4.775)(14.0,-5.975)
\psframe[linecolor=black, linewidth=0.04, dimen=outer](18.0,-4.775)(16.0,-5.975)
\psdots[linecolor=black, dotsize=0.1](0.4,-2.775)
\psdots[linecolor=black, dotsize=0.1](0.4,-3.175)
\rput[bl](1.6,-4.775){$\tilde{H}_{\tilde{M}-1}(z)$}
\rput[bl](16.4,-5.575){$\tilde{F}_{\tilde{M}-1}(z)$}
\rput[bl](14.0,-5.575){$\uparrow \frac{M_{\tilde{M}-1}}{2}$}
\rput[bl](3.6,-4.775){$\downarrow \frac{M_{\tilde{M}-1}}{2}$}
\rput[bl](6.4,-5.575){$\downarrow 2$}
\rput[bl](6.4,-4.375){$\downarrow 2$}
\rput[bl](8.4,-4.375){j.Im}
\rput[bl](8.4,-5.575){Re}
\rput[bl](12.4,-4.375){$\uparrow 2$}
\rput[bl](12.4,-5.575){$\uparrow 2$}
\psline[linecolor=black, linewidth=0.04, arrowsize=0.05291666666666667cm 2.0,arrowlength=1.4,arrowinset=0.0]{->}(0.4,-4.375)(1.2,-4.375)
\psline[linecolor=black, linewidth=0.04, arrowsize=0.05291666666666667cm 2.0,arrowlength=1.4,arrowinset=0.0]{->}(0.4,-1.175)(0.4,-2.375)
\psdots[linecolor=black, dotsize=0.1](18.8,-3.975)
\psline[linecolor=black, linewidth=0.04, arrowsize=0.05291666666666667cm 2.0,arrowlength=1.4,arrowinset=0.0]{->}(0.4,-3.575)(0.4,-4.375)
\psline[linecolor=black, linewidth=0.04, arrowsize=0.05291666666666667cm 2.0,arrowlength=1.4,arrowinset=0.0]{->}(18.8,-2.375)(18.8,-3.175)
\psdots[linecolor=black, dotsize=0.1](18.8,-3.575)
\psline[linecolor=black, linewidth=0.04, arrowsize=0.05291666666666667cm 2.0,arrowlength=1.4,arrowinset=0.0]{->}(11.2,-4.375)(12.0,-4.375)
\psline[linecolor=black, linewidth=0.04, arrowsize=0.05291666666666667cm 2.0,arrowlength=1.4,arrowinset=0.0]{->}(11.2,-5.575)(12.0,-5.575)
\psline[linecolor=black, linewidth=0.04](9.2,-4.375)(10.0,-4.375)
\psline[linecolor=black, linewidth=0.04](9.2,-5.575)(10.0,-5.575)
\psline[linecolor=black, linewidth=0.04, arrowsize=0.05291666666666667cm 2.0,arrowlength=1.4,arrowinset=0.0]{->}(5.2,-4.375)(6.0,-4.375)
\psline[linecolor=black, linewidth=0.04, arrowsize=0.05291666666666667cm 2.0,arrowlength=1.4,arrowinset=0.0]{->}(7.2,-4.375)(8.0,-4.375)
\psline[linecolor=black, linewidth=0.04, arrowsize=0.05291666666666667cm 2.0,arrowlength=1.4,arrowinset=0.0]{->}(7.2,-5.575)(8.0,-5.575)
\psline[linecolor=black, linewidth=0.04, arrowsize=0.05291666666666667cm 2.0,arrowlength=1.4,arrowinset=0.0]{->}(13.2,-5.575)(14.0,-5.575)
\psline[linecolor=black, linewidth=0.04, arrowsize=0.05291666666666667cm 2.0,arrowlength=1.4,arrowinset=0.0]{->}(15.6,-5.575)(16.0,-5.575)
\psline[linecolor=black, linewidth=0.04, arrowsize=0.05291666666666667cm 2.0,arrowlength=1.4,arrowinset=0.0]{->}(18.0,-5.575)(18.8,-5.575)
\psline[linecolor=black, linewidth=0.04, arrowsize=0.05291666666666667cm 2.0,arrowlength=1.4,arrowinset=0.0]{->}(18.8,-4.375)(18.8,-5.575)
\psline[linecolor=black, linewidth=0.04, arrowsize=0.05291666666666667cm 2.0,arrowlength=1.4,arrowinset=0.0]{->}(13.6,-4.375)(13.6,-5.575)
\psline[linecolor=black, linewidth=0.04](5.6,-4.375)(5.6,-5.575)
\psline[linecolor=black, linewidth=0.04, arrowsize=0.05291666666666667cm 2.0,arrowlength=1.4,arrowinset=0.0]{->}(5.6,-5.575)(6.0,-5.575)
\psline[linecolor=black, linewidth=0.04](13.2,-4.375)(13.6,-4.375)
\rput[bl](10.0,-4.375){$Y_{\tilde{M}-1}^{(I)}(z)$}
\rput[bl](10.0,-5.575){$Y_{\tilde{M}-1}^{(R)}(z)$}
\rput[bl](5.2,-5.975){$z^{-1}$}
\rput[bl](13.6,-4.375){$z^{-1}$}
\end{pspicture}
}

\caption{Non-uniform Modified DFT filter bank}
\label{non-uniform mdft}

\end{figure}

\par The non-uniform bands are obtained by merging the adjacent analysis and synthesis filters. Consider the analysis filter $\tilde{H}_{i}(z)$, which is obtained by merging $p_{i}$ adjacent analysis filters, as given below.

  \begin {equation}
\tilde{H}_{i}(z) = \sum_{k=n_{i}}^{n_{i}+p_{i} - 1}H_{k}(z) , \quad i= 0,1,......\tilde{M}-1
\end {equation}

where $n_{i}$ is the channel number and $p_{i}$ is the number of adjacent channels to be combined.

The synthesis filter $\tilde{F}_{i}(z)$, is obtained in a similar way.

 \begin {equation}
\tilde{F}_{i}(z) = \frac{1}{p_{i}}\sum_{k=n_{i}}^{n_{i}+p_{i} - 1}F_{k}(z) , \quad i= 0,1,......\tilde{M}-1
\end {equation}

\par The $\tilde{H}_{i}(z)$ and $\tilde{F}_{i}(z)$, $i= 0,1,......\tilde{M}-1$, form a new set of analysis and synthesis filters in the $\tilde{M}$-channel non-uniform MDFT FB. The corresponding decimation factor $M_{i}$, is given by $M_{i}$ = $\frac{M}{p_{i}}$. 

\par In order for such a non-uniform MDFT FB to become a valid FB, each analysis filter $\mid \tilde{H}_{i}(e ^{j\omega})\mid$ should have a band selection capability. For this we assume that the prototype filter $H(z)$ is originally designed to meet, in addition to bandlimit condition, the flatness condition 

\begin {equation}
\begin {aligned}
\mid H(e^{j\omega})\mid ^ {2} + \mid H(e^{j(\omega-\frac{2\pi}{M})})\mid^{2} = 1, \\
 for \quad  0 \leq \omega \leq \frac{2\pi}{M}
 \end {aligned}
 \end {equation}

Then we obtain the following property:

\par
Property 1: The analysis filter $\tilde{H}_{i}(z)$ and the synthesis filter $\tilde{F}_{i}(z)$, $i= 0,1,......\tilde{M}-1$, satisfy the flatness condition within the pass-band in $(n_{i} \frac{2\pi}{M}+ \varepsilon,(n_{i}+p_{i}) \frac{2\pi}{M}-\varepsilon)$ and $((-n_{i}+p_{i}) \frac{2\pi}{M}+\varepsilon,-n_{i} \frac{2\pi}{M}- \varepsilon)$

\par Proof: Here it is verified only for the analysis filters $\tilde{H}_{i}(z)$. Similarly it can be proved for the synthesis filter $\tilde{F}_{i}(z)$. The condition can be written as \citep{lee1995design} 

\begin {equation}
\begin {aligned}
\mid \tilde{H}_{i}(e^{j\omega}) \mid ^ {2}  &= \sum_{k=n_{i}}^{n_{i}+p_{i} - 1} \mid H_{k}(e^{j\omega}) \mid ^ {2} \\
 &\quad + \sum_{k=n_{i}}^{n_{i}+p_{i} - 2} [ H_{k}(e^{j\omega}) H^{\ast}_{k+1}(e^{j\omega}) \\ 
 &\quad + H^{\ast}_{k}(e^{j\omega}) H_{k+1}(e^{j\omega})]
\end {aligned}
\end {equation}

where $H^{\ast}_{k}(e^{j\omega})$ is the conjugate complex of $H_{k}(e^{j\omega})$ and  $H^{\ast}_{k+1}(e^{j\omega})$ is the conjugate complex of $H_{k+1}(e^{j\omega})$.

\par The adjacent aliasing components in equation (11) are eliminated in MDFT FB structure \citep{fliege1994modified}. Then the equation (11) can be written as
\begin {equation}
\mid \tilde{H}_{i}(e^{j\omega}) \mid ^ {2}  =  \sum_{k=n_{i}}^{n_{i}+p_{i} - 1} \mid H_{k}(e^{j(\omega-k\frac{2\pi}{M})})\mid ^ {2}  
\end {equation}

Hence the analysis filter $\tilde{H}_{i}(z)$ meets the flatness condition. 

\par From equation (6), it is proved that all adjacent aliasing terms and all odd aliasing terms are compensated in uniform MDFT FB \citep{fliege1994modified,karp1999modified}. The remaining alias spectra are removed by an adequate high stopband attenuation of the synthesis filter. 

Property 2: A condition is derived to find a channel which will lead to aliasing, when it is merged with any combination of adjacent channels. 

The derived design validity is verified using the alias cancellation among the non-uniform analysis and synthesis filters, by giving a sinusoidal input to the filter bank with all even values of M \citep{fliege1994modified}.

\par The channel number which leads to aliasing when merged with adjacent channels is experimentally obtained as $a_{M}$.  

		\begin {equation}
a_{M} = (\frac{M-2n+2}{2})-1    
\end {equation}
		
	where $M$ is the total number of channels and $n$ is an integer derived as given below.

	If $M = 10i+4$ or $M= 10i+6$ for $i=0,1,2,3.....$ , then
	
	\begin {equation}
	n= \left\lfloor \frac{M+1}{5}\right\rfloor 
	\end {equation}
	
	\par If $M = 10i+8$ or $M= 10i+10$ or $M= 10i+12$  for $i=0,1,2,3.....$ , then
	
	\begin {equation}
	n= \left\lfloor \frac{M+2}{5}\right\rfloor 
	\end {equation}
	   
	$M$ is always chosen as even, since a decimation of $M/2$ is required in the structure of MDFT filter banks. The $n$ value obtained from equation (14) or (15) when substituted in equation (13) gives the channel number $a_{M}$. For example, for M=8, equation (15) is to be used to obtain the value of $n$ as 2. Hence from equation (13), $a_{M}$ is obtained as 2. This means the channel number 2 if merged with any other adjacent channels will lead to aliasing. For any value of M, only a single channel when merged with other adjacent channels results in aliasing. Hence for $M=8$, merging of channels 0 and 1, 3 and 4, 4 and 5, 5 and 6, 6 and 7, 3,4 and 5, 4,5 and 6, 5,6 and 7, 3,4,5 and 6, 4,5,6 and 7 and 3,4,5,6 and 7 is possible without causing aliasing.

Property 3: The distortion function of a non-uniform MDFT FB is similar to uniform MDFT FB.

The distortion function can be written as \citep{lee1995design} 
			
			\begin {equation}
\tilde{T}_{dist}(z) = \frac{1}{M}\sum_{i=0}^{\tilde{M} - 1}\sum_{k=n_{i}}^{n_{i}+p_{i} - 1}\sum_{m=n_{i}}^{n_{i}+p_{i} - 1}F_{k}(z)H_{m}(z) 
\end {equation}

\par Due to the bandlimit condition of the prototype filter, only the adjacent terms will remain in equation (16). Hence we can rewrite equation (16) as

\begin {equation}
\begin {aligned}
\tilde{T}_{dist}(z) &= \frac{1}{M}\sum_{i=0}^{\tilde{M} - 1}\sum_{k=n_{i}}^{n_{i}+p_{i} - 1}F_{k}(z)H_{k}(z)\\
 &\quad + \frac{1}{M}\sum_{i=0}^{\tilde{M} - 1}\sum_{k=n_{i}}^{n_{i}+p_{i} - 2}(F_{k}(z)H_{k+1}(z) + F_{k+1}(z)H_{k}(z))
\end {aligned}
\end {equation}

The adjacent aliasing terms are canceled within the MDFT FB \citep{fliege1994modified} and by inserting the equations (1) and (2) in (17), the second term in equation (17) becomes zero. Hence  equation (17) can be written as

\begin {equation}
\tilde{T}_{dist}(z) = \frac{1}{M}\sum_{i=0}^{\tilde{M} - 1}\sum_{k=n_{i}}^{n_{i}+p_{i} - 1}F_{k}(z)H_{k}(z) 
\end {equation}

Therefore the distortion functions of uniform and non-uniform MDFT FB are found to be same.

\par The first property verifies that associated adjacent analysis (or synthesis) filters produce filter which meet the flatness condition. Second property identifies conditions of aliasing in non-uniform MDFT FB. From this condition, we can get the alias-free M-channel non-uniform filters. Third property confirms that the distortion function is the same as that of uniform MDFT FB.

In the MDFT FB, the linear phase prototype finite impulse response(FIR) filter designed using Parks McClellan method, is complex modulated to derive all the analysis and synthesis filters with linear phase. Based on the degree of overlap between the adjacent filter responses, some amplitude distortion will occur. In this paper, the passband and stopband edge frequencies are iteratively adjusted, with fixed transition width to satisfy the 3-dB condition \citep{kalathil2014non}.\\{Design Specifications}:\\Maximum pass-band ripple :$0.004dB$\\Minimum stop-band attenuation :$60dB$\\Pass-band edge frequency :$0.0618\pi$\\Stop-band edge frequency :$0.0634\pi$\\Number of channels :$8$\\

Initially an 8-channel uniform Modified DFT FB is designed using Parks McClellan method. By appropriately merging the filters of 8 channel uniform MDFT FB, the non-uniform FB are designed. The merging of channels is done for all possible and valid combinations of adjacent channels. To verify the adjacent channel aliasing during merging of channels, a sinusoidal signal is given as input to the MDFT FB and the output is verified for each combination of merging channels. Fig. 4 shows the frequency response plots of the analysis filters of 8-channel uniform , non-uniform 5-channel derived from 8-channel uniform MDFT FB and non-uniform 4-channel derived from 8-channel uniform FB . The frequency response plots of the analysis filters of 16-channel uniform , non-uniform 8-channel derived from 16-channel uniform MDFT FB and non-uniform 6-channel derived from 16-channel uniform MDFT FB are shown in Fig. 5. The input and output spectrum of the non-uniform MDFT FB are shown in Fig. 6. Amplitude distortion function plot for the 4 and 6-channel non-uniform MDFT FB is shown in Fig. 7. Two illustrations are shown to find the aliased channel for any number of channels.

\begin{figure}[here]
\centering
\subfloat[]{\includegraphics[width=2.5in]{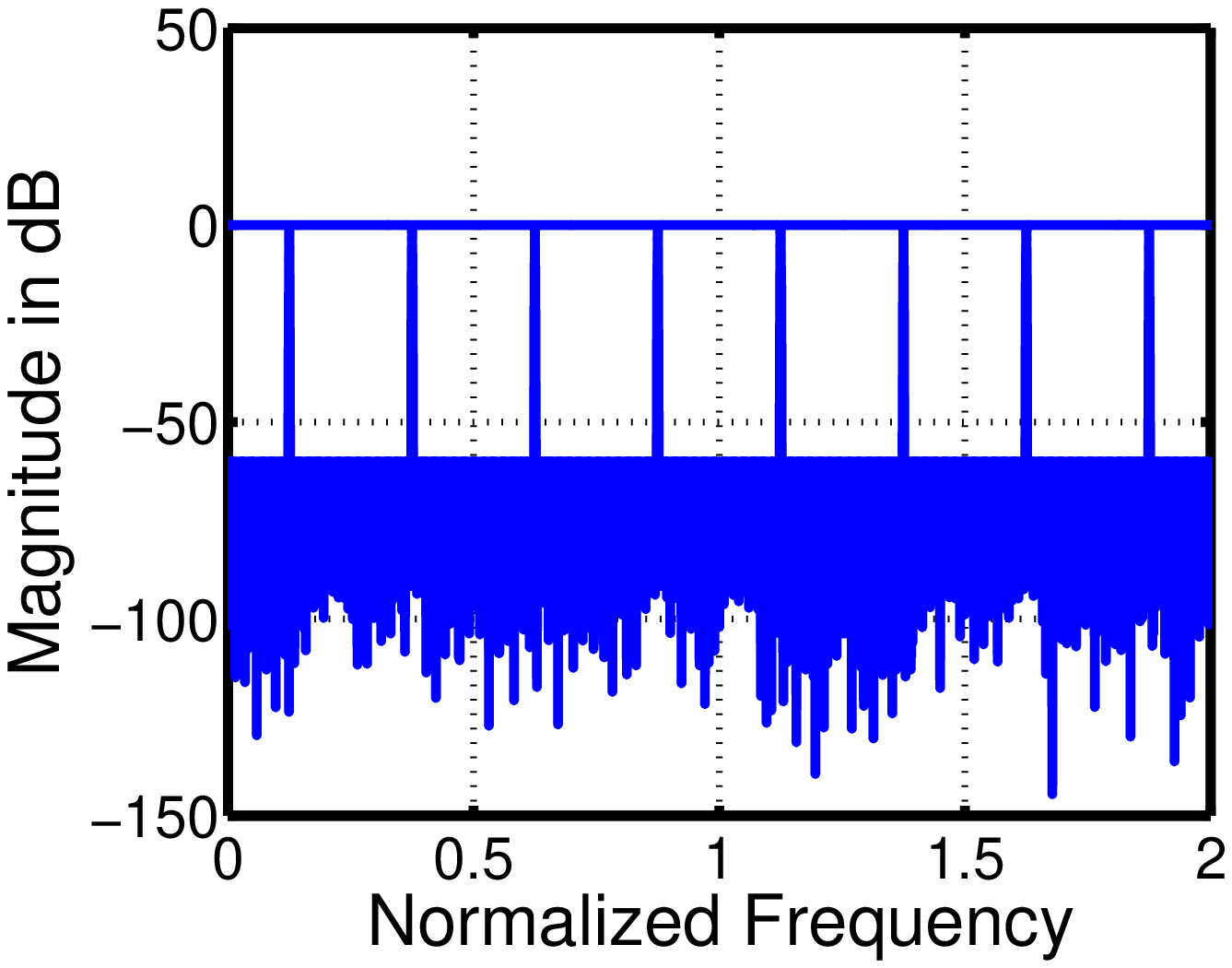}
\label{fig_first_case}}
\hfil
\subfloat[]{\includegraphics[width=2.5in]{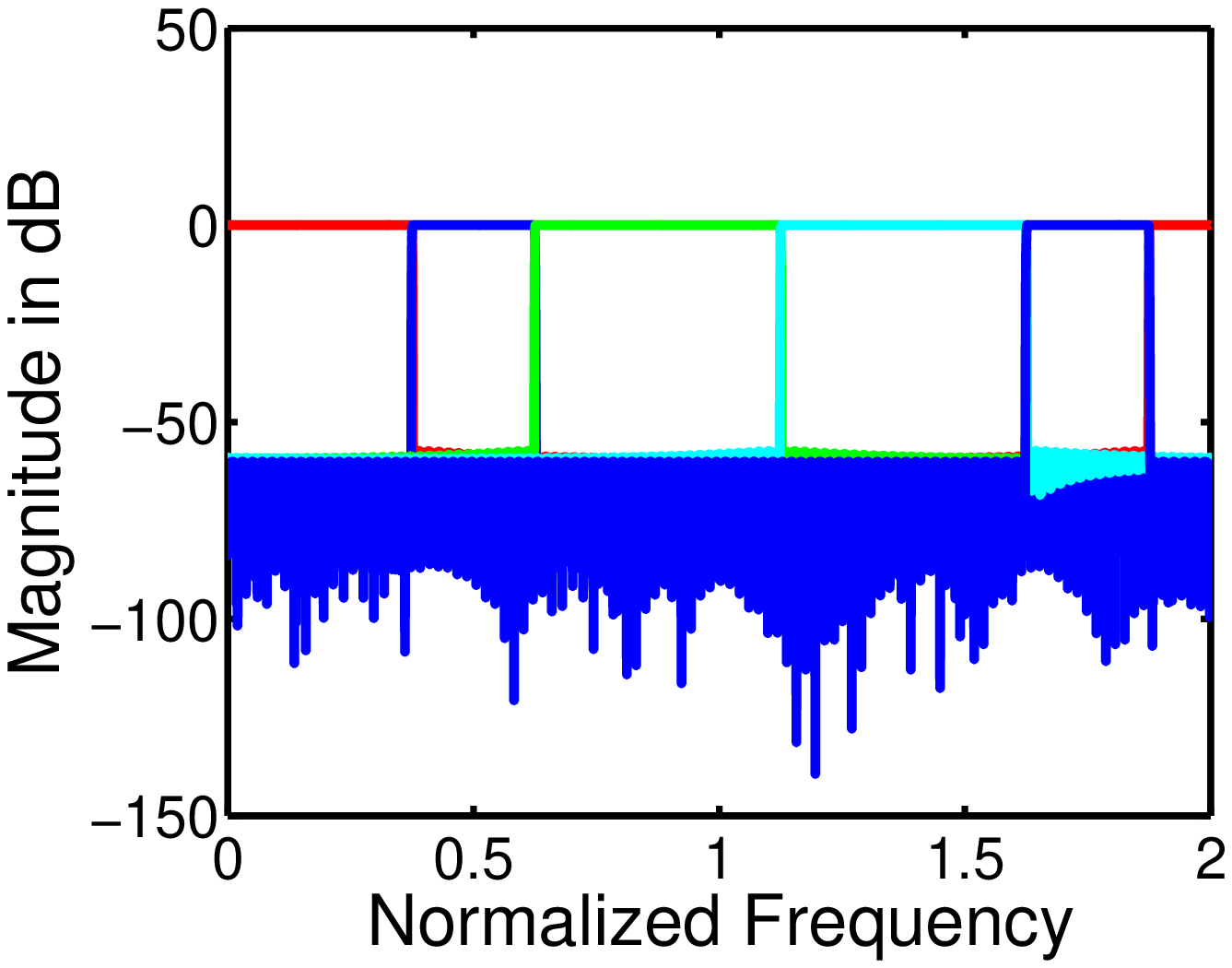}
\label{fig_second_case}}
\hfil
\subfloat[]{\includegraphics[width=2.5in]{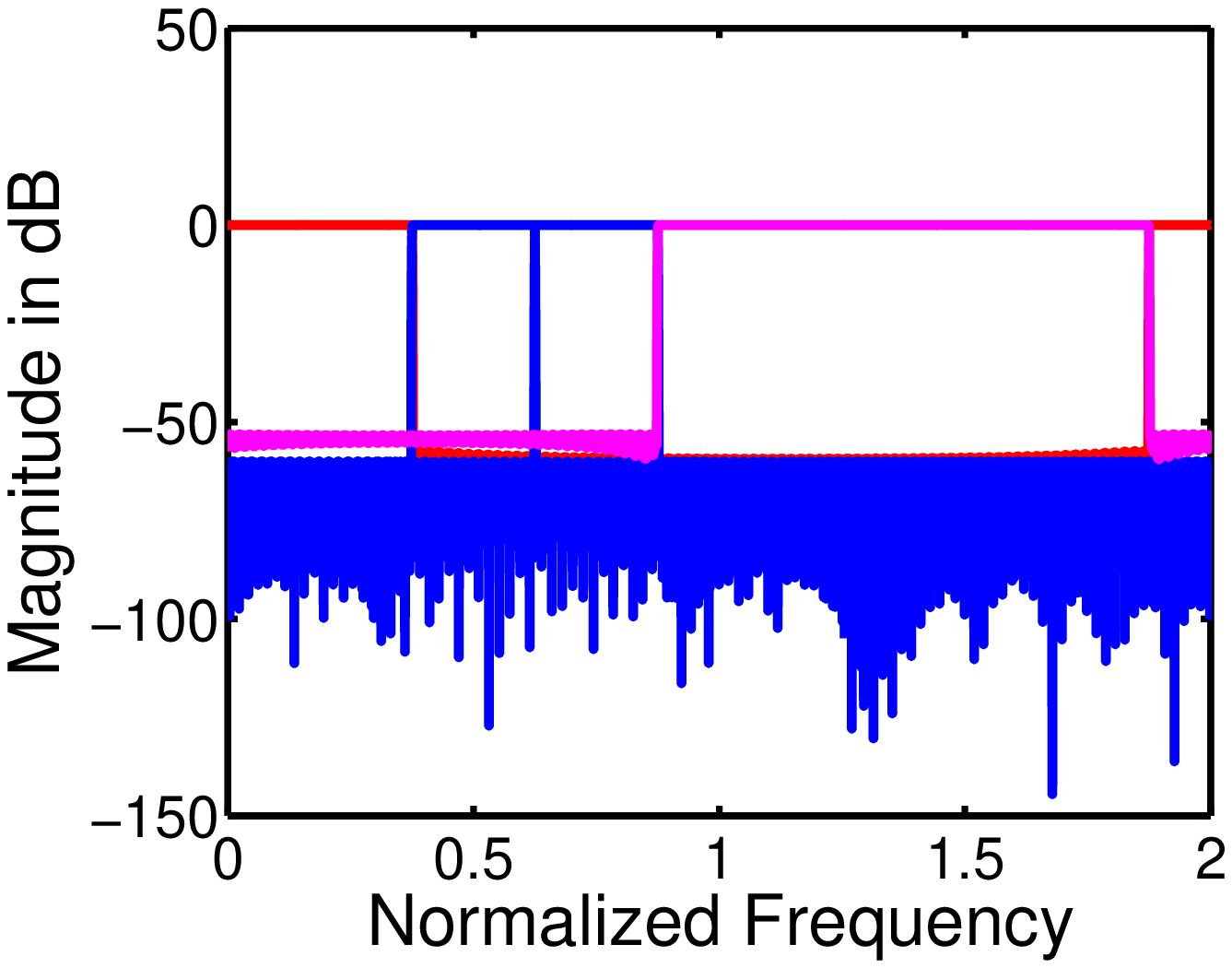}
\label{fig_third_case}}
\caption{Frequency Response of analysis filters (a) 8-channel uniform MDFT filter bank (b) Non-uniform 5-channel filter bank ( 4, 8, 4, 4, 8 ) (c) Non-uniform 4-channel filter bank (4, 8, 8, 2)}
\label{fig_sim}
\end{figure}

  \begin{figure}[here]
\centering
\subfloat[]{\includegraphics[width=2.5in]{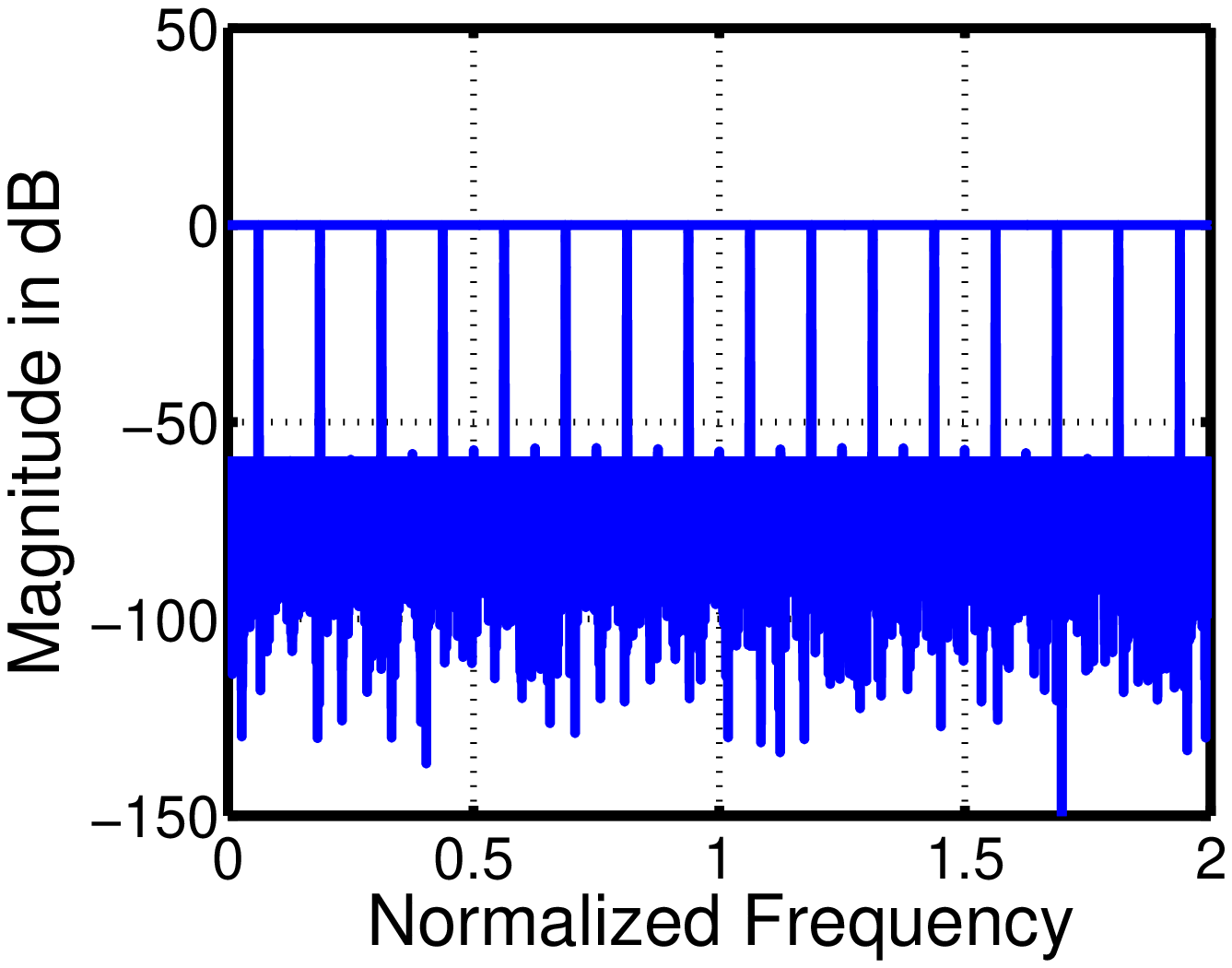}
\label{fig_first_case}}
\hfil
\subfloat[]{\includegraphics[width=2.5in]{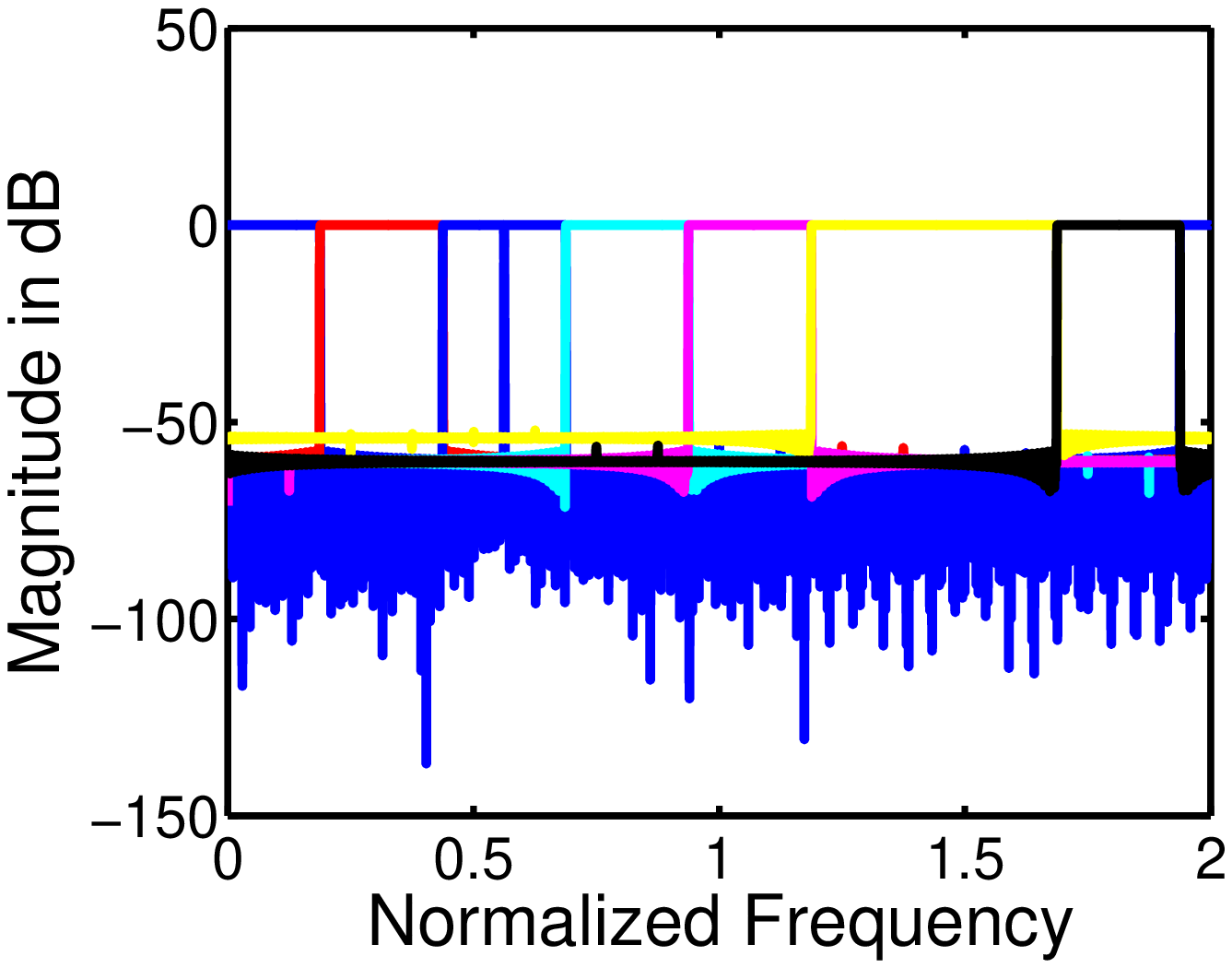}
\label{fig_first_case}}
\hfil
\subfloat[]{\includegraphics[width=2.5in]{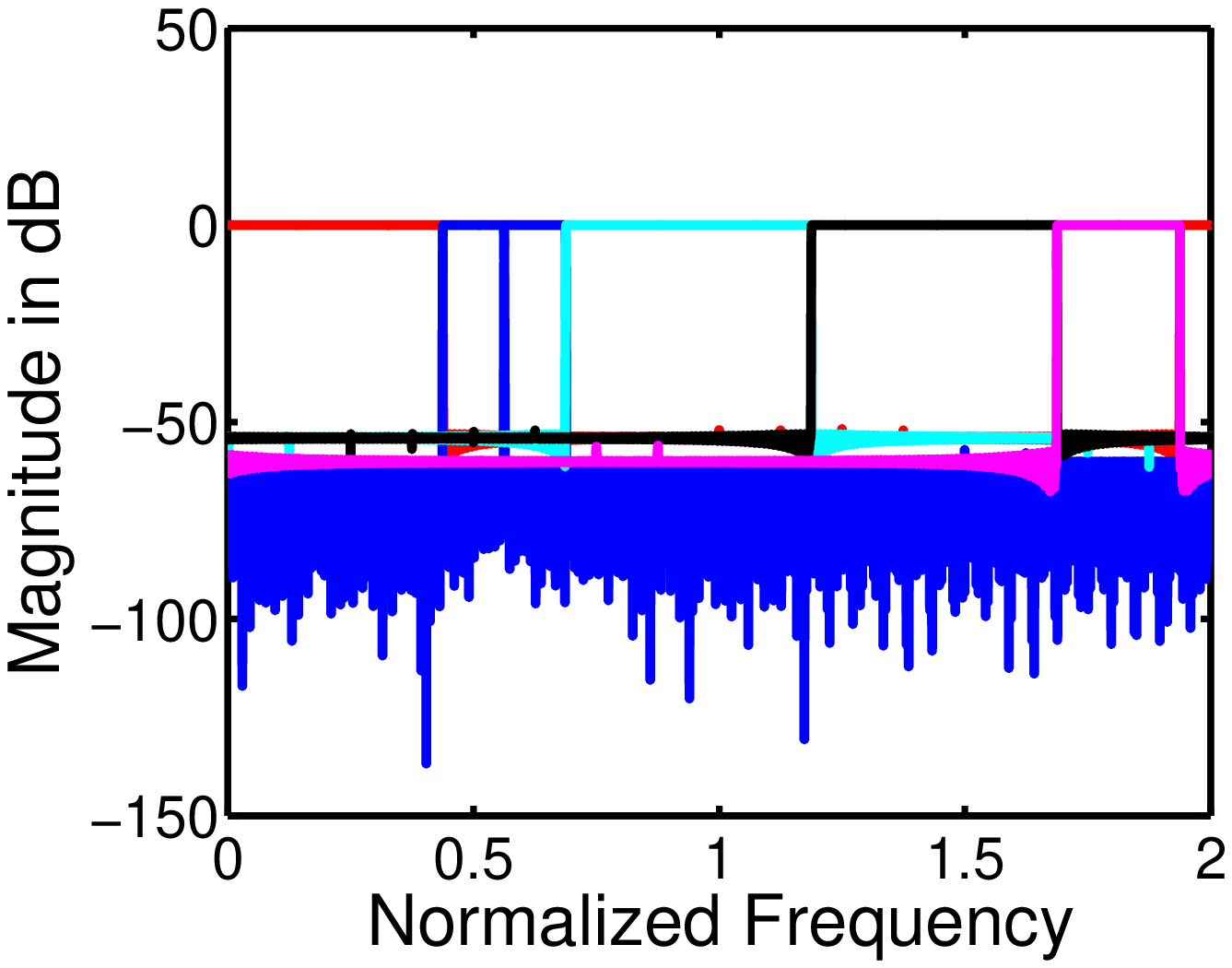}
\label{fig_second_case}}
\caption{ Frequency Response of analysis filters (a) 16-channel uniform MDFT filter bank (b) Non-uniform 8-channel filter bank ( 8, 8, 16, 16, 8, 8, 4, 8 ) (c) Non-uniform 6-channel filter bank ( 4, 16, 16, 4, 4, 8 )}
\label{fig_sim}
\end{figure}

\begin{figure}[here]
\centering
\subfloat[]{\includegraphics[width=2.5in]{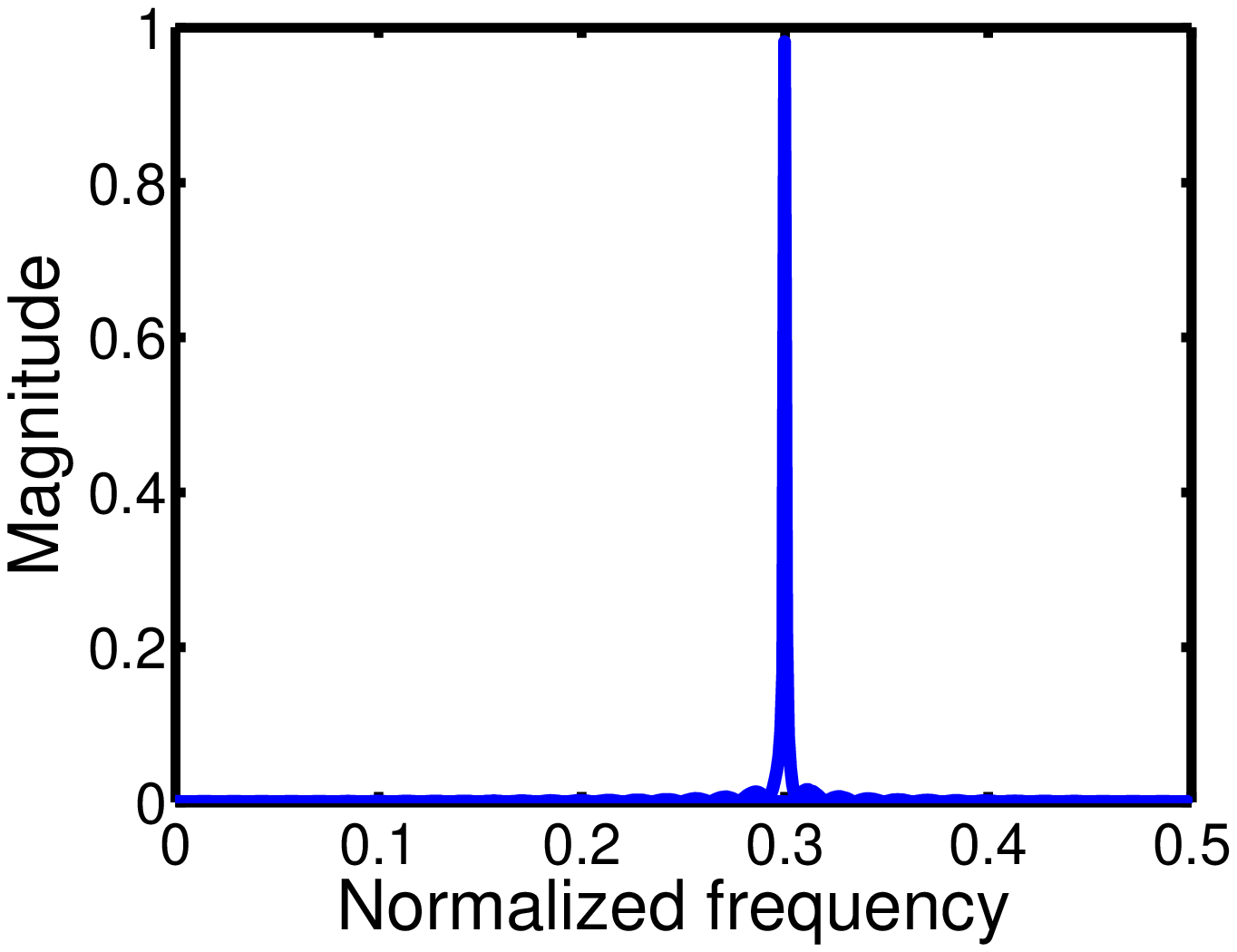}
\label{fig_first_case}}
\hfil
\subfloat[]{\includegraphics[width=2.5in]{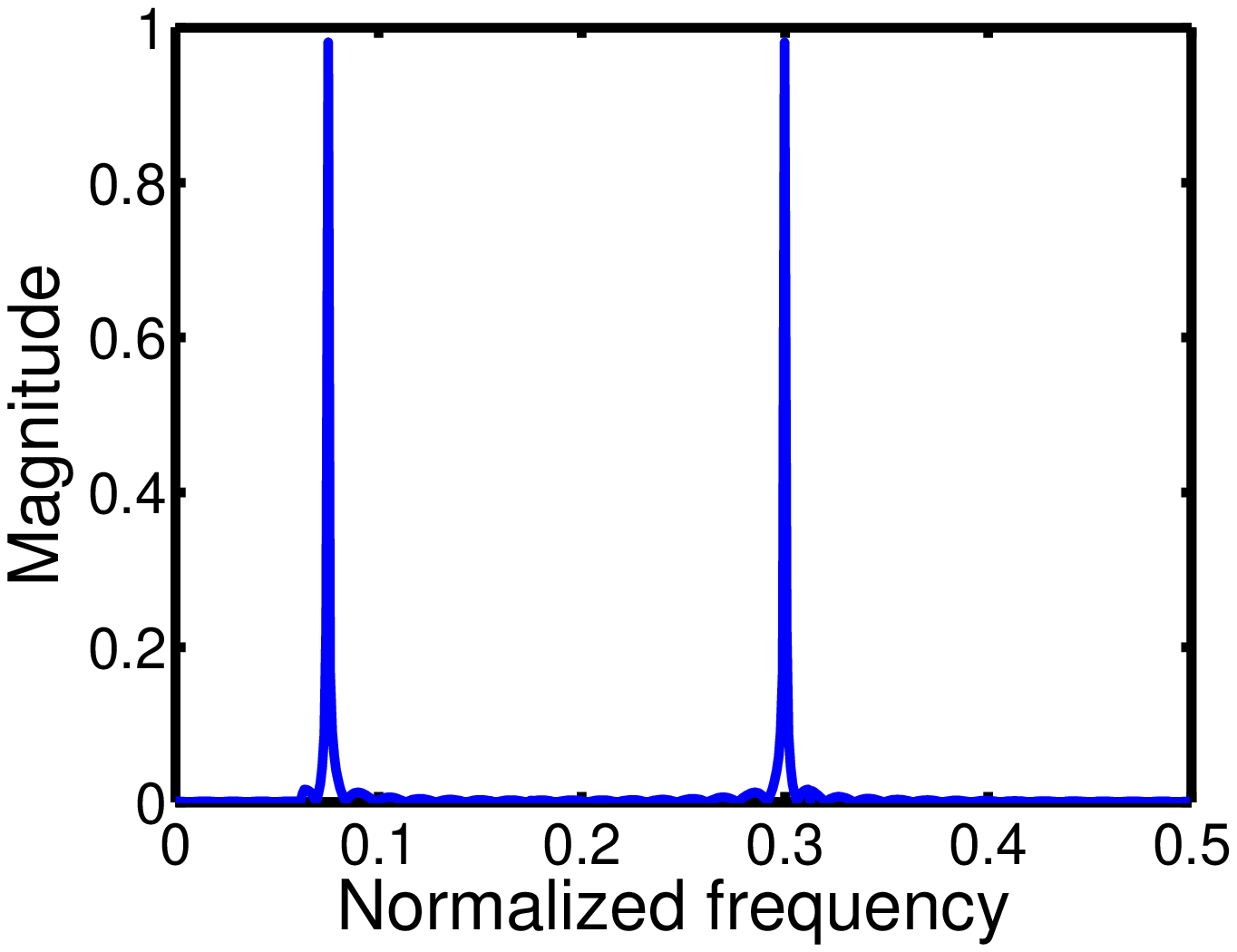}
\label{fig_second_case}}
\hfil
\subfloat[]{\includegraphics[width=2.5in]{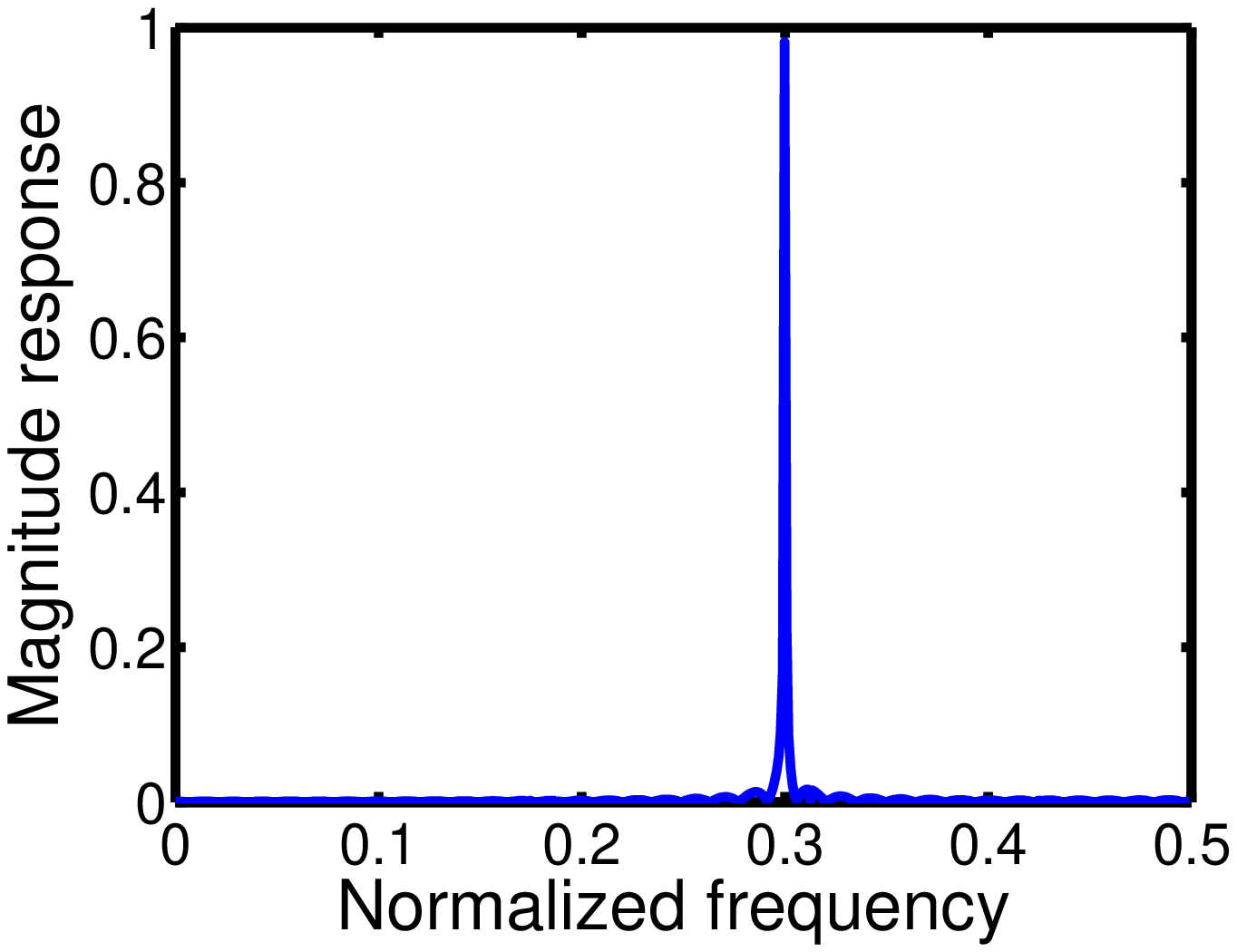}
\label{fig_third_case}}

\caption{(a) Input signal spectrum. (b) Aliasing in output spectrum after merging channels 2 and 3 in 8- channel uniform filter bank (c) No aliasing in output spectrum after merging channels 4 and 5 in 8- channel uniform filter bank}
\label{fig_sim}
\end{figure}

\begin{figure}[here]
\centering
\subfloat[]{\includegraphics[width=2.5in]{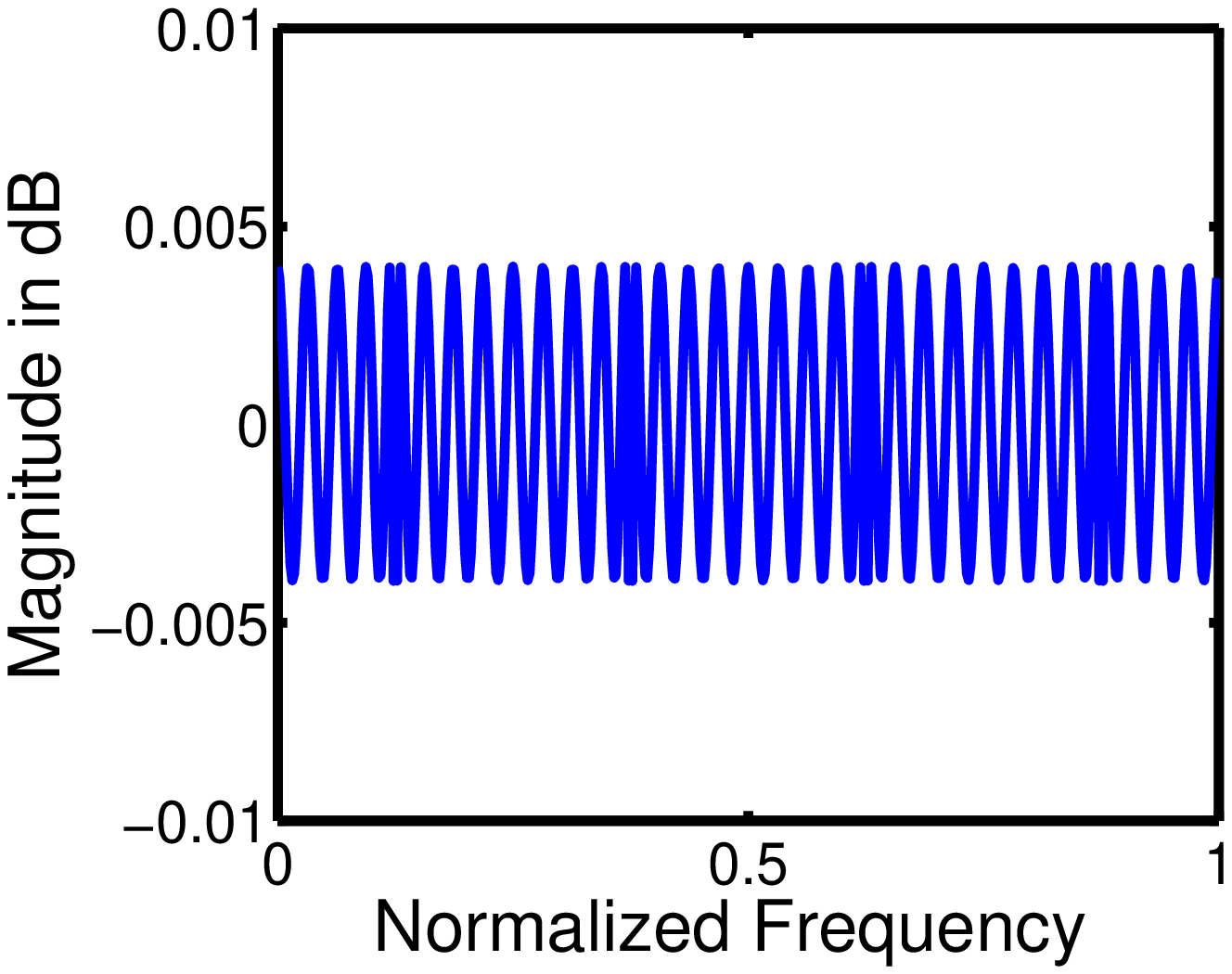}
\label{fig_first_case}}
\hfil
\subfloat[]{\includegraphics[width=2.5in]{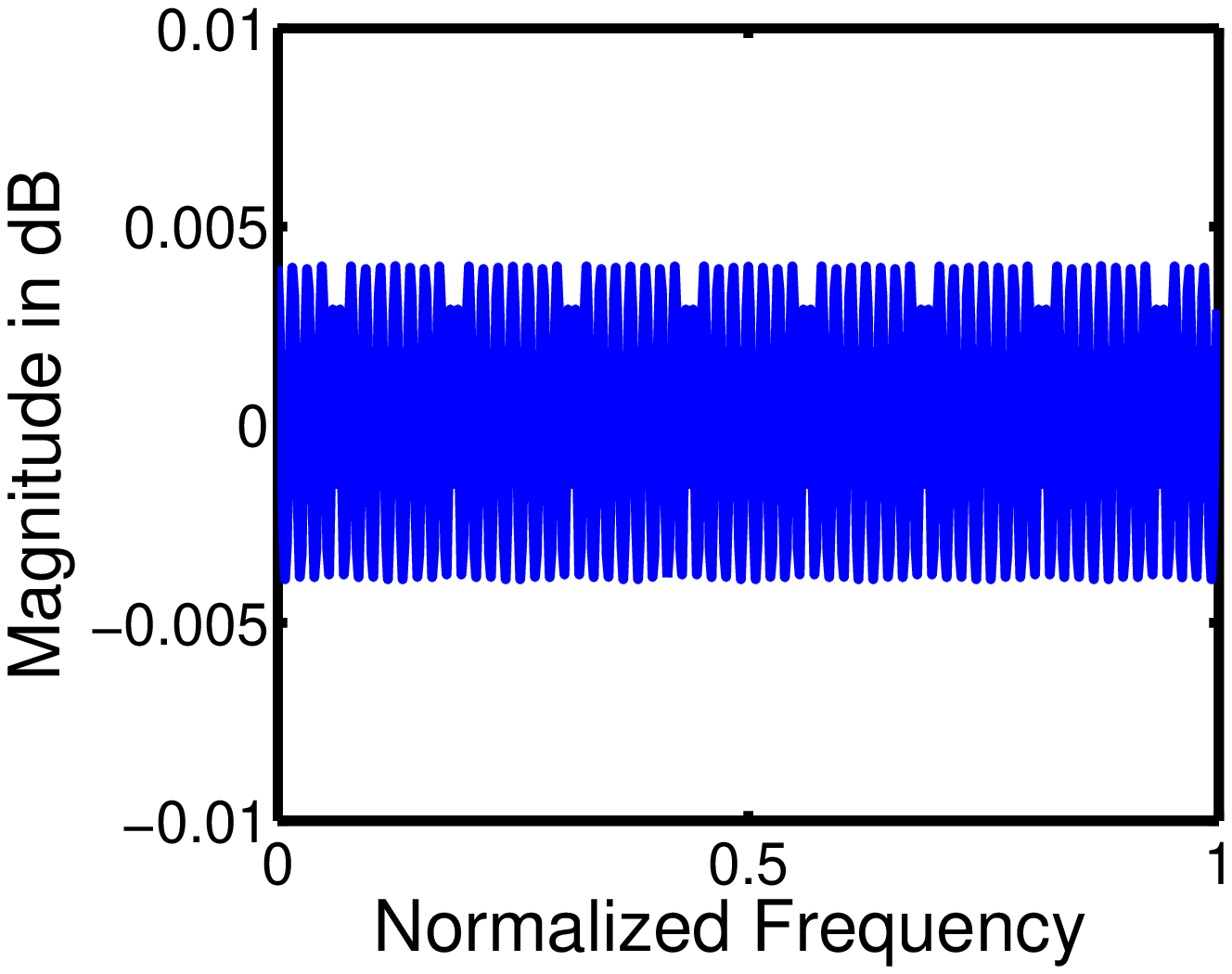}
\label{fig_second_case}}

\caption{Amplitude distortion function plot of non-uniform MDFT filter bank (a) 4-channel (b) 6-channel }
\label{fig_sim}
\end{figure}

Illustration 1

If $M$=8, then it will come under the constraint $M= 10i+8$. Here $i=0$. Hence equation (15) is to be used to obtain $n$. 

 $n= \left\lfloor \frac{M+2}{5}\right\rfloor = 2$

	then $a_{M} = (\frac{M-2n+2}{2})-1 = 2$

		This shows that if channel 2 is merged with any combination of adjacent channels it will lead to aliasing during merging.

Illustration 2

If $M$=14, $M$ will come under the constraint $M= 10i+4 ; i=1$. Then $n$ is obtained from equation (14) as 3 and $a_{14}=4$.

		This shows that channel 4 and its combinations will give aliasing during merging.
   
\section{Results and Discussion }

	Initially, for the given specifications, the prototype filter with linear phase is designed for uniform MDFT FB. Using the uniform filters, the non-uniform MDFT filters are designed by merging adjacent channels. The adjacent channels to be merged are selected making use of equations (13), (14) and (15). All the simulations are performed on an Intel(R) core(TM) i5 processor operating at 2.4 GHz using MATLAB 7.10.0(R2010a). The design of non-uniform CMFB using merging of adjacent channels is already available \citep{lee1995design,kalathil2014non}. But in the non-uniform CMFB , the constituent filters do not have linear phase even with a linear phase prototype filter. In the non-uniform MDFT FB proposed in this paper, all the constituent filters have linear phase.

\section{Conclusion}

Non-uniform FB satisfying linear phase property for all the constituent filters, are desirable in applications such as speech and image processing and in communication. The design of non-uniform MDFT FB derived from a uniform MDFT FB by merging appropriate adjacent channels is proposed in this paper. The proposed design of the non-uniform MDFT FB is checked utilizing the alias cancellation among the channels, distortion and flatness condition of the channels. An equation is derived to find the condition to prevent aliasing. Since the structure of the uniform MDFT FB is preserved in the non-uniform FB, the phase distortion is zero and amplitude distortion is kept very small.



\bibliographystyle{model5-names}\biboptions{authoryear}
\bibliography{refer}





\end{document}